# Decomposition of multi-particle azimuthal correlations in Q-cumulant analysis

L. Nađđerđ[1,2], J. Milošević[1,2], D. Devetak[1,2], and F. Wang[3,2,4]
[1]*Department of physics, "VINČA" Institute of Nuclear Science - National Institute of the Republic of Serbia, University of Belgrade, Mike Petrovića Alasa 12-14, Vinča 11351, Belgrade, Serbia.*
[2]*Strong coupling Physics International Research Laboratory, Huzhou University, Huzhou, Zhejiang 313000, P. R. China.*
[3]*College of Science, Huzhou University, Huzhou, Zhejiang 313000, P. R. China.*
[4]*Department of Physics and Astronomy, Purdue University, Indiana 47907, USA.*

**Abstract**

The method of Q-cumulants is a powerful tool to study the fine details of azimuthal anisotropies in high energy nuclear collisions. This paper presents a new method, based on mathematical induction, to evaluate the analytical form of the high-order Q-cumulants. The ability of this method is demonstrated via a toy model that uses the elliptic power distribution to simulate anisotropic emission of particles, quantified in terms of Fourier flow harmonics $v_n$ . The method can help in studying the large amount of event statistics that can be collected in the future, and to allow measurements of very high central moments of the $v_2$ distribution. This can in turn facilitate progress in understanding the initial geometry, input to hydrodynamic calculations of medium expansion in high energy nuclear collisions, as well as constraints on it.

## 1 Introduction

In ultra-relativistic collisions of nuclei at both the Relativistic Heavy Ion Collider (RHIC) [1-4] and the Large Hadron Collider (LHC) [5-8] a hot and dense system of strongly interacting quarks and gluons called the "quark-gluon plasma" (QGP) is created. One of the key observables used to study properties of the QGP is the anisotropic collective flow that is quantified by Fourier harmonics $v_n$. Many methods have been developed to measure these harmonics [9-11]. One of them, the method of cumulants based on multi-particle correlations [12, 13], enables to suppress short-range correlations arising from jets and resonance decays and then reveals a genuine collectivity arising from the expansion of the QGP. The Q-cumulant method [14] is an improved version of the initial cumulant method. Recently, the Q-cumulants have been employed to examine hydrodynamics predictions in Ref. [15] based on the ratio of the differences between adjacent cumulants [16]. This analysis, performed using the CMS data, indicated the necessity of introducing moments higher than skewness to describe finer details of the elliptic flow $v_2$ distribution. In order to do so, higher-order cumulants have to be calculated. In this paper we present a way how to determine expressions needed to calculate higher-order cumulants.

The paper is organized as follows. Section 2 gives the basic quantities used in the Q-cumulant method. Section 3 gives a foundation of the method within a finite-dimensional vector space. Section 4 gives the results together with applications of the method to simple cases. Using toy models, we demonstrate the capabilities of the method in Section 5. A summary is given in Section 6. Tables that summarize coefficients needed for expressing the cumulants up to the 14-th order can be found in the Appendix.

## 2 Basics of the Q-cumulants

The general formalism of cumulants, for the purpose of flow measurements, was first used in Refs. [12-14]. Cumulants were expressed in terms of moments of the magnitude of the corresponding flow vector. Such cumulant method is systematically biased due to trivial effects of same particle autocorrelations. To avoid the computation expensive "nested loops" in removing autocorrelations in the measurements of multi-particle azimuthal correlations, Refs. [12-14] have proposed an alternative method, which is based on the use of generating functions. An improved cumulant method, the Q-cumulants [14] allows, in principle, for a fast and exact calculation of all multi-particle cumulants. However, in practice the determination of the analytical expressions of multi-particle cumulants, which involve azimuthal correlations more than eight particles, is very difficult.

The Q-cumulants are built upon analytical expressions for multi-particle azimuthal correlations in terms of flow vectors $Q$, with:

$$Q_n = \sum_{j=1}^{M} \mathrm{e}^{in\phi_j} \;, \tag{1}$$

evaluated in different orders of Fourier harmonics, $n$. Here $M$ is the multiplicity or number of particles in an event, and $\phi_j$ labels the azimuthal angle of the $j$-th particle measured in a fixed coordinate system in the laboratory. The method involves a decomposition of $2m$-th power of the magnitude of the flow vector, $|Q_n|^{2m}$:

$$\left|Q_n\right|^{2m} = \sum_{j_1,\ldots,j_{2m}=1}^{M} \mathrm{e}^{in\left(\phi_{j_1}+\ldots+\phi_{j_m}-\phi_{j_{m+1}}-\ldots-\phi_{j_{2m}}\right)} , \tag{2}$$

into (off diagonal) terms having $2m$, $2m$-1, … different indices up to the term where all $2m$ indices are equal (diagonal term). The first term, having $2m$ different indices, of decomposition is proportional to the $2m$-particle azimuthal correlations $\langle 2m \rangle$:

$$\sum_{j_1\neq\ldots\neq j_{2m}=1}^{M} \mathrm{e}^{in\left(\phi_{j_1}+\ldots+\phi_{j_m}-\phi_{j_{m+1}}-\ldots-\phi_{j_{2m}}\right)} \equiv P_{M,2m} \cdot \langle 2m \rangle , \tag{3}$$

where $P_{M,2m}$ is the number of distinct $2m$-particle combinations one can form for an event with multiplicity $M$:

$$P_{M,2m} = \frac{M!}{(M-2m)!} \,. \tag{4}$$

In the case of full decomposition, $|Q_n|^{2m}$ is expressed in readily calculable terms of powers of the flow vector given with Eq. (1) along with the anticipated $2m$-particle azimuthal correlations introduced in Eq. (3). A thorough derivation with detailed examples is presented in Ref. [17]. The analytical decomposition of $|Q_n|^{2m}$ for $m > 4$, however, becomes tedious.

In this paper we present a numerical method of decomposition of $|Q_n|^{2m}$ that, with the use of modern computers, enables one to easily obtain analytical expressions of multi-particle azimuthal correlations of higher orders.

## 3 Foundation of the method

The full decomposition of $|Q_n|^{2m}$ consists in partition of appropriate sets (or multisets in the case of repeated elements) of the azimuthal angles that lead to the partial Bell polynomials or their generalization (case of multisets). In derivation, some parts of the Bell polynomials have to be expressed through the lower-order multi-particle azimuthal correlations. This change of side in the expression makes a sign change of the corresponding term, resulting in coefficients of the partial Bell polynomials in the final form of the $2m$-particle azimuthal correlations to be positive or negative integers. For example, in the case of $m = 2$ one obtains the four-particle azimuthal correlations might be presented as a linear combination of the corresponding Q-vector terms:

$$P_{M,4}\langle 4\rangle = +1|Q_n|^4 - 2\,\mathrm{Re}\left[Q_{2n}Q_n^*Q_n^*\right] + 1|Q_{2n}|^2 - 4(M-2)|Q_n|^2 + 2M(M-3), \qquad (5)$$

with the corresponding integer coefficients (+1, -2, +1, -4, +2) in front of each of the terms.

The same holds true for all $2m$-particle correlations but with different sets of integer coefficients. This fact inspired us to calculate all these integer coefficients by solving an appropriate system of algebraic equations.

The $2m$-particle correlations $\langle 2m\rangle$ might be considered as a member of a finite-dimensional vector space $V_d$, where $d = \frac{1}{2}\sum_{l=0}^{m} p(l)[p(l)+1]$ is a dimension of the vector space, and $p(l) = \{1, 1, 2, 5, 7, \ldots\}$ is a partition function of a non-negative integer $l = \{0, 1, 2, 3, 4, \ldots\}$, respectively. In $V_d$ one may define a basis $B_{\langle 2m\rangle} = (\mathbf{e}_1, \mathbf{e}_2, \mathbf{e}_3, \ldots, \mathbf{e}_d)$ that enables one to present $P_{M,2m}\langle 2m\rangle$ as a linear combination of basis vectors $\mathbf{e}_i$:

$$P_{M,2m}\langle 2m\rangle = x_1 f_1^{(m,l)}\mathbf{e}_1 + x_2 f_2^{(m,l)}\mathbf{e}_2 + \ldots + x_d f_d^{(m,l)}\mathbf{e}_d\,, \qquad (6)$$

where $x_1$, $x_2$, … are unknown integer numbers that need to be obtained and $f_i^{(m,l)}$ are known integer functions of multiplicity $M$. The left side of Eq. (6) can be calculated directly from Eq. (3) only in case of low multiplicity, being computationally manageable. The $d$ unknown integer numbers, $x_1, x_2, \ldots, x_d$ on the right hand side of the Eq. (6) might be calculated numerically by solving a system of $d$ linear algebraic equations presented in the next Section.

## 4 Determination of the basis $B_{\langle 2m\rangle} = (\mathbf{e}_1, \mathbf{e}_2, \mathbf{e}_3, \ldots, \mathbf{e}_d)$

The full analytical determination of the basis $B_{\langle 2m\rangle}$ is presented in Ref. [17] for $m = 1,\ldots,4$. However, in this paper we show straightforward determination of the basis by the method of mathematical induction, which enables easy calculation of the multi-particle azimuthal correlations of higher orders. First, it should be noticed, by a simple inspection of the published results [14, 17], that each basis $B_{\langle 2m\rangle}$ contains the complete basis of the lower number (2*m*-2)-particle azimuthal correlations $B_{\langle 2m-2\rangle} \subset B_{\langle 2m\rangle}$. For example, the basis of 4-particle azimuthal correlations contains the complete basis of the 2-particle azimuthal correlations $B_{\langle 2\rangle} = \left(1, |Q_n|^2\right)$ and some additional basis vectors:

$$B_{\langle 4\rangle} = \left(B_{\langle 2\rangle}, |Q_n|^4, \mathrm{Re}(Q_{2n}Q_n^*Q_n^*), |Q_{2n}|^2\right). \tag{7}$$

Also, the basis $B_{\langle 2\rangle} = \left(B_{\langle 0\rangle}, |Q_n|^2\right)$ might be considered as an extension of the basis $B_{\langle 0\rangle}$, with an additional basis vector $|Q_n|^2$. All additional vectors in a basis $B_{\langle 2l\rangle}$ form a subset *l*, whose dimension is $d(l) = \frac{1}{2} p(l)\left[p(l)+1\right]$. Thus, the problem of determining the basis $B_{\langle 2m\rangle}$ reduces to finding the subset *l* = *m*.

We found that the subset *l* = *m* contains different compositions (products) of the flow vectors $Q_{sub1}Q_{sub2}\cdots$ each having a subscript (*sub*1, *sub*2,…) that corresponds to the well-known "partition of the positive integer" *m*. For example, in the case of the 8-particle (2*m* = 8) azimuthal correlations, the subset *l* = 4 contains these compositions of the flow vectors:

$$\begin{matrix} Q_{1n}Q_{1n}Q_{1n}Q_{1n} \\ Q_{2n}Q_{1n}Q_{1n} \\ Q_{2n}Q_{2n} \\ Q_{3n}Q_{1n} \\ Q_{4n} \end{matrix} \quad \Leftarrow \quad \begin{cases} 1 \quad 1 \quad 1 \quad 1 \\ 2 \quad 1 \quad 1 \\ 2 \quad 2 \\ 3 \quad 1 \\ 4 \end{cases}, \tag{8}$$

where each composition contains subscripts that correspond to the partition of the integer 4.

Here we have used a convenient symbolic notation introduced in Refs. [14, 17]:

$$Q_{p\cdot n} \equiv \sum_{j=1}^{M} e^{p\cdot in\phi_j}, \; p \in \{1,2,3,\ldots\}. \tag{9}$$

To obtain all basis vectors of the subset $l$ = 4, each composition in (8) has to be combined by the ordered composition of the complex conjugate vectors as is shown in the following pattern:

$$
\begin{array}{lcl}
Q_nQ_nQ_nQ_n & \Leftarrow & Q_n^*Q_n^*Q_n^*Q_n^* \\
Q_{2n}Q_nQ_n & \Leftarrow & Q_{2n}^*Q_n^*Q_n^* \\
Q_{2n}Q_{2n} & \Leftarrow & Q_{2n}^*Q_{2n}^* \\
Q_{3n}Q_n & \Leftarrow & Q_{3n}^*Q_n^* \\
Q_{4n} & \Leftarrow & Q_{4n}^*
\end{array}
\qquad (10)
$$

This pattern gives fifteen (p(4)[p(4)+1]/2 = 15) different combinations:

$$
\begin{array}{lll}
Q_nQ_nQ_nQ_n \,|\, Q_n^*Q_n^*Q_n^*Q_n^* & Q_{2n}Q_nQ_n \,|\, Q_n^*Q_n^*Q_n^*Q_n^* & Q_{2n}Q_{2n} \,|\, Q_n^*Q_n^*Q_n^*Q_n^* \\
 & Q_{2n}Q_nQ_n \,|\, Q_{2n}^*Q_n^*Q_n^* & Q_{2n}Q_{2n} \,|\, Q_{2n}^*Q_n^*Q_n^* \\
 & & Q_{2n}Q_{2n} \,|\, Q_{2n}^*Q_{2n}^*
\end{array}
$$

$$
\begin{array}{ll}
Q_{3n}Q_n \,|\, Q_n^*Q_n^*Q_n^*Q_n^* & Q_{4n} \,|\, Q_n^*Q_n^*Q_n^*Q_n^* \\
Q_{3n}Q_n \,|\, Q_{2n}^*Q_n^*Q_n^* & Q_{4n} \,|\, Q_{2n}^*Q_n^*Q_n^* \\
Q_{3n}Q_n \,|\, Q_{2n}^*Q_{2n}^* & Q_{4n} \,|\, Q_{2n}^*Q_{2n}^* \\
Q_{3n}Q_n \,|\, Q_{3n}^*Q_n^* & Q_{4n} \,|\, Q_{3n}^*Q_n^* \\
 & Q_{4n} \,|\, Q_{4n}^*
\end{array}
\qquad (11)
$$

The real part of each of these additional vectors (11), together with the basis $B_{\langle 6\rangle}$, makes the complete basis of the 8-particle azimuthal correlations $B_{\langle 8\rangle}$. This presentation of the composition of the flow vectors that are separated from the complex conjugate part by a vertical bar is inspired by the work of Ref. [14].

The corresponding multiplying integer functions $f^{(m,l)}$ might be also obtained via mathematical induction. These integer functions are given by:

$$
\begin{array}{rll}
(M-2l)\prod_{j=m+l+1}^{2m-1}(M-j), & \text{for} & l=\{0,...,m-2\} \\
M-2l, & \text{for} & l=m-1 \\
1, & \text{for} & l=m
\end{array}
. \qquad (12)
$$

For example, in the case of 8-particle azimuthal correlations ($m$ = 4):

$$
f^{(m=4,l)} = \begin{cases}
M(M-5)(M-6)(M-7), & \text{for} \quad l=0 \\
(M-2)(M-6)(M-7), & \text{for} \quad l=1 \\
(M-4)(M-7), & \text{for} \quad l=2 \\
(M-6), & \text{for} \quad l=3 \\
1, & \text{for} \quad l=4
\end{cases}. \qquad (13)
$$

We list these integer functions and basis vectors for each of the 2*m*-particle azimuthal correlations ( $m = 1,2,\ldots,7$ ) in the third and fourth columns respectively of the corresponding Table 1-7.

Now we have all necessary quantities to form the system of linear equations to obtain all the unknown coefficients for each of the 2*m*-particle azimuthal correlations:

$$\begin{aligned}
\left(\sum_{j_1\neq\ldots\neq j_{2m}=1}^{M} \mathrm{e}^{in\left(\phi_{j_1}+\ldots+\phi_{j_m}-\phi_{j_{m+1}}-\ldots-\phi_{j_{2m}}\right)}\right)_{\text{event1}} &= x_1\left(f_1^{(m,l)}\mathbf{e}_1\right)_{\text{event1}} + x_2\left(f_2^{(m,l)}\mathbf{e}_2\right)_{\text{event1}} + \ldots + x_d\left(f_d^{(m,l)}\mathbf{e}_d\right)_{\text{event1}} \\
\left(\sum_{j_1\neq\ldots\neq j_{2m}=1}^{M} \mathrm{e}^{in\left(\phi_{j_1}+\ldots+\phi_{j_m}-\phi_{j_{m+1}}-\ldots-\phi_{j_{2m}}\right)}\right)_{\text{event2}} &= x_1\left(f_1^{(m,l)}\mathbf{e}_1\right)_{\text{event2}} + x_2\left(f_2^{(m,l)}\mathbf{e}_2\right)_{\text{event2}} + \ldots + x_d\left(f_d^{(m,l)}\mathbf{e}_d\right)_{\text{event2}} \\
\vdots \qquad &= \qquad \vdots \\
\left(\sum_{j_1\neq\ldots\neq j_{2m}=1}^{M} \mathrm{e}^{in\left(\phi_{j_1}+\ldots+\phi_{j_m}-\phi_{j_{m+1}}-\ldots-\phi_{j_{2m}}\right)}\right)_{\text{event}d} &= x_1\left(f_1^{(m,l)}\mathbf{e}_1\right)_{\text{event}d} + x_2\left(f_2^{(m,l)}\mathbf{e}_2\right)_{\text{event}d} + \ldots + x_d\left(f_d^{(m,l)}\mathbf{e}_d\right)_{\text{event}d}
\end{aligned} \quad ,(14)$$

In order to set a solvable system of equations (14) the multiplicity has to be $M \geq 2m$ (otherwise, the system of equations cannot be formed). For example, the system of equations for the 2-particle azimuthal correlations is given by ( $M \geq 2$ ):

$$\begin{aligned}
\left(\sum_{j_1\neq j_2=1}^{M} \mathrm{e}^{in\left(\phi_{j_1}-\phi_{j_2}\right)}\right)_{\text{event1}} &= x_1\left(M\cdot 1\right)_{\text{event1}} + x_2\left(1\cdot\left|Q_n\right|^2\right)_{\text{event1}} \\
\left(\sum_{j_1\neq j_2=1}^{M} \mathrm{e}^{in\left(\phi_{j_1}-\phi_{j_2}\right)}\right)_{\text{event2}} &= x_1\left(M\cdot 1\right)_{\text{event2}} + x_2\left(1\cdot\left|Q_n\right|^2\right)_{\text{event2}}
\end{aligned} \quad . \qquad (15)$$

One can notice that in the sum in the left-handed side of the Eq. (15) are, in fact, added a complex number and its conjugate, so the sum will be reduced to a real number. In order to set the system of equations (15) we randomly chose two sets of angles: $\phi_{j_1} = 0.400906$ , $\phi_{j_2} = -2.84149$ , $\phi_{j_3} = 1.98067$ (multiplicity *M* = 3) for the first equation and $\phi_{j_1} = -1.32161$ , $\phi_{j_2} = 2.75646$ , $\phi_{j_3} = 2.8089$ , $\phi_{j_4} = 1.59479$ , $\phi_{j_5} = 1.13565$ , (*M* = 5) for the second one. Then, the system of equations given by Eq. (15) for *n* = 2 becomes:

$$\begin{aligned}
-1.99218068\,2386965 &= x_1\cdot 3 + x_2\cdot 1.00781931\,7613035 \\
-2.80896273\,559164 &= x_1\cdot 5 + x_2\cdot 2.19103726\,440836
\end{aligned} \quad . \qquad (16)$$

By using the same sets of azimuthal angles but now for Fourier harmonic *n* = 3, the system of equations (15) becomes:

$$\begin{aligned}
-2.50222960\,4410126 &= x_1\cdot 3 + x_2\cdot 0.49777039\,5589874 \\
2.85297668\,985644 &= x_1\cdot 5 + x_2\cdot 7.85297668\,985644
\end{aligned} \quad . \qquad (17)$$

These two systems of linear equations are mutually different but both have the same solution: $x_1 = -1,\ x_2 = 1$ so we fill in the second column of the Table 1 with the corresponding $x_i$ values.

The sets of the azimuthal angles might be chosen randomly, as we did in the upper example. That should secure that the calculated basis $B_{\langle 2m \rangle}$ do not have collinear vectors, otherwise the linear system of equations is not solvable.

The calculated numbers that enter the system of Eqs. (16, 17) should be given with appropriate number of significant digits in order to get the obtained coefficients $x_i$ as true integers. There are no criteria to determine the appropriate number of significant digits. One can use the 'trial and error' method to find out the required number. Otherwise, the obtained $x_i$ will not be integers and the rounding of $x_i$ might be problematic in the case of large basis $B_{\langle 2m \rangle}$. The limits of the application of this method depend on the number of significant digits by which the computer operates. We have successfully applied the method up to the 14-th order of cumulants.

**Table 1** Coefficients, integer functions and basis vectors for calculation of the two-particle azimuthal correlations.

| $i$ | $x_i$ | $f_i^{(m=1,l)}$ | **Basis vectors** | $l$ |
|---|---|---|---|---|
| 1 | -1 | $M$ | 1 | 0 |
| 2 | 1 | 1 | $Q_n \mid Q_n^*$ | 1 |

The two-particle azimuthal correlation formula can be formed by simply reading the corresponding values from Table 1:

$$\langle 2 \rangle P_{M,2} = x_1 f_1^{(m=1,0)} 1 + x_2 f_2^{(m=1,1)} \operatorname{Re}\left(Q_n Q_n^*\right) = -M + \left|Q_n\right|^2 . \tag{18}$$

The similar table can be obtained for the four-particle azimuthal correlations by the solution of the corresponding linear system of equations:

**Table 2** Coefficients, integer functions and basis vectors for calculation of the four-particle azimuthal correlations.

| $i$ | $x_i$ | $f_i^{(m=2,l)}$ | **Basis vectors** | $l$ |
|---|---|---|---|---|
| 1 | 2 | $M(M-3)$ | 1 | 0 |
| 2 | -4 | $(M-2)$ | $Q_n \mid Q_n^*$ | 1 |
| 3 | 1 | 1 | $Q_n Q_n \mid Q_n^* Q_n^*$ | 2 |
| 4 | -2 | 1 | $Q_{2n} \mid Q_n^* Q_n^*$ | 2 |
| 5 | 1 | 1 | $Q_{2n} \mid Q_{2n}^*$ | 2 |

That might be also easily rewritten in an appropriate formula by simply reading the corresponding values of Table 2:

$$\begin{aligned} \langle 4 \rangle P_{M,4} &= x_1 f_1^{(m=2,0)} 1 + x_2 f_2^{(m=2,1)} \operatorname{Re}(Q_n Q_n^*) + x_3 f_3^{(m=2,2)} \operatorname{Re}(Q_n Q_n Q_n^* Q_n^*) + \\ &+ x_4 f_4^{(m=2,2)} \operatorname{Re}(Q_{2n} Q_n^* Q_n^*) + x_5 f_5^{(m=2,2)} \operatorname{Re}(Q_{2n} Q_{2n}^*) \\ &= 2M(M-3) - 4(M-2)\left|Q_n\right|^2 + \left|Q_n\right|^4 - 2\operatorname{Re}(Q_{2n} Q_n^* Q_n^*) + \left|Q_{2n}\right|^2 \end{aligned} . \tag{19}$$

The appropriate tables for the higher order cumulants are given in the Appendix.

Many of the basis vectors are complex valued numbers like: $Q_{2n}Q_n^*Q_n^*$ and one should take only their real part when one writes the expression of the $2m$-particle azimuthal correlations. Actually, the analytical derivation reveals that in addition to these complex valued basis vectors their complex conjugates ($Q_{2n}^*Q_nQ_n$) also participate in the corresponding basis. Because of symmetry, they always enter the expression of the $2m$-particle azimuthal correlations with the same coefficients and so their imaginary parts cancel. This also causes the corresponding coefficients to be even numbers. For example, $x_i' f_i^{(m,l)} Q_{2n}Q_n^*Q_n^* + x_i' f_i^{(m,l)} Q_{2n}^*Q_nQ_n = 2x_i' f_i^{(m,l)} \operatorname{Re}(Q_{2n}Q_n^*Q_n^*)$, and so: $x_i = 2x_i'$ is always even number. Odd coefficients might appear only in front of basis vectors that are not accompanied with their complex conjugates like $|Q_n|^4$ and $|Q_{2n}|^2$ in Eq. (19).

The coefficients have also other interesting features that should be noticed. For example, all coefficients which correspond to the same subset of basis vectors (having the same index $l$) sum up to zero: (Table 3.: $\sum_{i=6}^{11} x_i = 0$, $\sum_{i=3}^{5} x_i = 0$), (Table 4.: $\sum_{i=12}^{26} x_i = 0$, and so on). The exceptions of this rule are the two coefficients at beginning of each Table that correspond to the basis vectors 1 and $|Q_n|^2$. The coefficient in front of the basis vector "1" is $(-1)^m m!$. One more interesting feature that might be used to check the correctness of the obtained expression of the $2m$-particle azimuthal correlations is a sum of the absolute values of all coefficients in a Table. This sum reads:

$$\sum_{i=1}^{d} |x_i| = \left\{ \begin{array}{rll} 2, & \text{for} & m=1 \\ 10, & \text{for} & m=2 \\ 96, & \text{for} & m=3 \\ 1560, & \text{for} & m=4 \\ 39120, & \text{for} & m=5 \\ 1409040, & \text{for} & m=6 \\ 69048000, & \text{for} & m=7 \end{array} \right\} = m!e\Gamma(m+1,1)\,, \tag{20}$$

where $d$ is the dimension of the vector space, $e$ is Euler's number, and $\Gamma(a,x) = \int_x^\infty t^{a-1}e^{-t}dt$ is the incomplete gamma function of $x$ with parameter $a$.

We also applied this method of derivation for the case of the weighted Q-vector evaluated in the harmonic $n$:

$$Q_{n,q} = \sum_{j=1}^{M} \omega_j^q \mathrm{e}^{in\phi_j}\,, \tag{21}$$

where $w_j$ is the weight of the $j$-th particle [17]. The derivation in this case is more complicated primarily because there are more basis vectors involved. However, the obtained coefficients in front of the basis vectors have very simple features:

$$\sum_{i=1}^{d'} x_i = 0\,, \tag{22}$$

$$\sum_{i=1}^{d'} |x_i| = (2m)!\,. \tag{23}$$

(The dimension of the vector space in this case is not equal to the previous one: $d' \neq d$ .)
Finally, having expressions for the $2m$-particle azimuthal correlations, $\langle 2m \rangle$ one can calculate the weighted average over all events $\langle\langle 2m \rangle\rangle$ (given by Eq. (1) in Ref. [18]). Then, one can use the recurrence relation to calculate the cumulants of any order by knowing all the cumulants of lower orders [18]:

$$c_n\{2k\} = \langle\langle 2k \rangle\rangle - \sum_{m=1}^{k-1} \binom{k}{m} \binom{k-1}{m} \langle\langle 2m \rangle\rangle c_n\{2k-2m\}. \tag{24}$$

The cumulant based flow harmonics $v_n\{2k\}$ ($k$ = 1, 2…) can be calculated by the following equation [16]:

$$v_n\{2k\} = \sqrt[2k]{a_{2k}^{-1} c_n\{2k\}}, \tag{25}$$

where coefficients $a_{2k}$ are obtainable by recursion relation:

$$a_{2k} = 1 - \sum_{m=1}^{k-1} \binom{k}{m} \binom{k-1}{m} a_{2k-2m}\ \text{, with: } a_2 = 1, \tag{26}$$

which enables easy calculation of high-order $v_n\{2k\}$ by using any commercial program for calculation. For example, $v_n\{2\} = \sqrt[2]{c_n\{2\}}$, $v_n\{8\} = \sqrt[8]{[-33]^{-1} c_n\{8\}}$,

$v_n\{16\} = \sqrt[16]{[\text{-}10643745]^{-1} c_n\{16\}}$, $v_n\{26\} = \sqrt[26]{[2473000014\ 7369440]^{-1} c_n\{26\}}$,

$v_n\{38\} = \sqrt[38]{[7069675533\ 2327402640\ 8967101760]^{-1} c_n\{38\}}$.

## 5 Demonstration of the method using a toy model

Experimental values of the $v_2\{2k\}$ enable determination of the central moments of the $v_2$ distribution. As a way to obtain the lowest central moments of the $v_2$ distribution as are variances $\sigma_{20}^2$ and $\sigma_{02}^2$, skewness $s_{30}$, and co-skewness $s_{12}$ [15, 16], it requires experimental values of at least four different cumulants $v_2\{2\}$, $v_2\{4\}$, $v_2\{6\}$, and $v_2\{8\}$. However, the centrality dependence of the hydrodynamics probe $h$ = ($v_2\{6\}$- $v_2\{8\}$)/( $v_2\{4\}$- $v_2\{6\}$) obtained in the experiment [16], indicates the non-zero value of the kurtosis $\kappa_{40}$, which is the fourth central moment of the $v_2$ distribution. So, to obtain the additional fourth central moments, $\kappa_{40}$, $\kappa_{22}$, $\kappa_{04}$ [16], it requires experimental values of at least three more cumulants of higher order $v_2\{10\}$, $v_2\{12\}$, $v_2\{14\}$.

Therefore, in order to show the validity of the above described method, the obtained expressions for the cumulants up to the 14-th order are calculated with the azimuthal angles simulated using toy models. For each event of a given $v_2$ a simple distribution $p(\phi) = \frac{1}{2\pi}\left(1 + 2v_2 \cos(2\phi)\right)$ is used to generate particle azimuthal angle. We set the input value $v_2 = 0.15$ with no event-by-event flow fluctuations. Thus, one expects that all $v_2\{2k\}$ are equal to the input value. Fig. 1 shows that this is indeed what is seen as there is excellent agreement between the cumulant based $v_2\{2k\}$ estimations and the input $v_2$ value of 0.15. Due to a strong correlation between different $\langle\langle 2m \rangle\rangle$ [18], the statistical uncertainties of the $v_2\{2k\}$ very slowly increases with the increase of $k$. This toy model therefore validates the method.

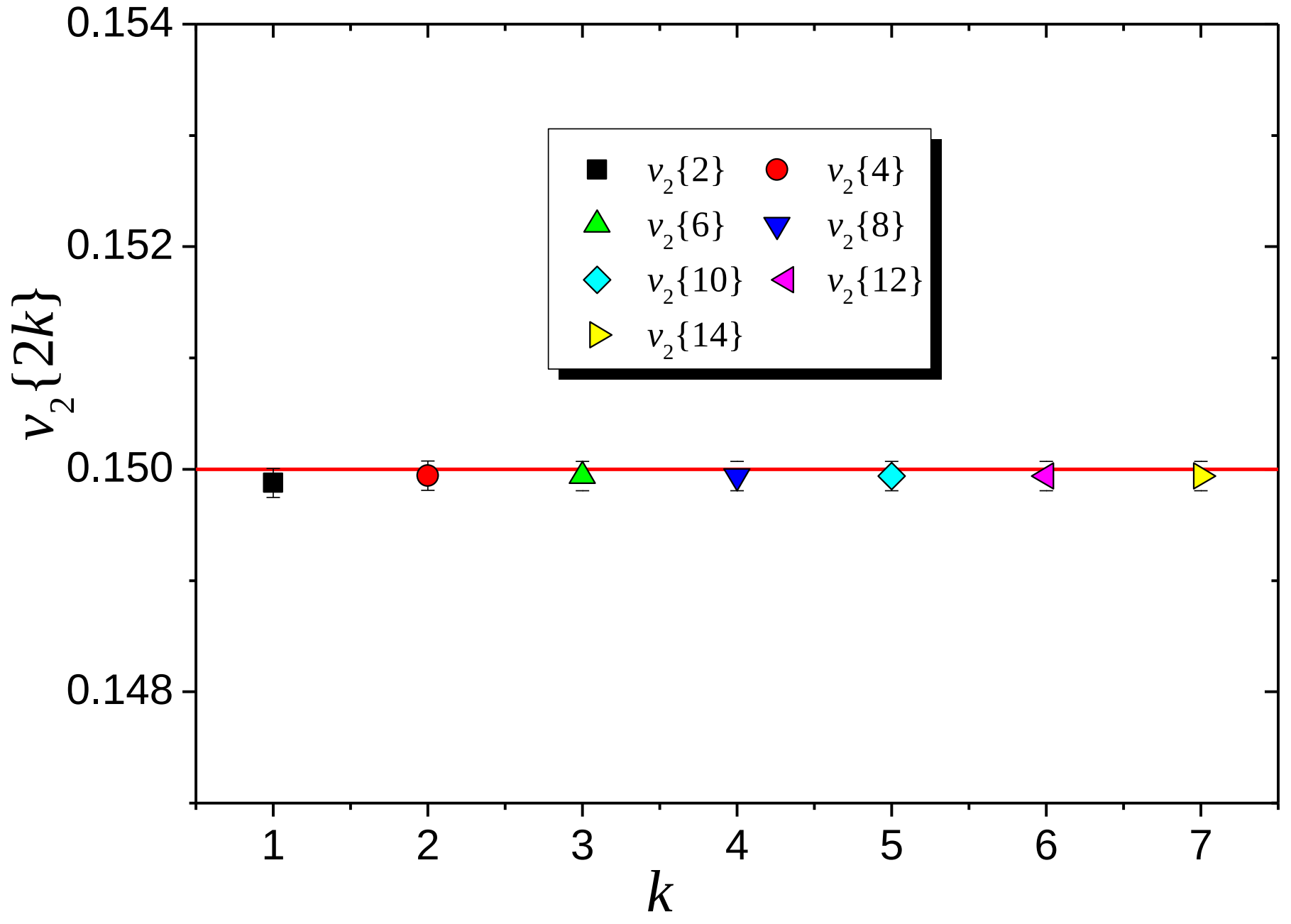

FIG. 1 Cumulant based estimations of the elliptic flow $v_2$ from toy Monte Carlo simulations. The input value of $v_2$ is 0.15. All cumulants retrieve this input value with high precision.

However, from this validation the sensitivity to flow fluctuations of the higher-order cumulants cannot be verified. Therefore, we applied the second toy model where the initial eccentricity $\varepsilon_2$ distribution is simulated using the elliptic power distribution:

$$\frac{dN}{d\varepsilon_2} = \frac{2\alpha}{\pi}(1-\varepsilon_0^2)^{\alpha+1/2}\varepsilon_2(1-\varepsilon_2^2)^{\alpha-1}\int_0^{\pi}\frac{1}{(1-\varepsilon_0\varepsilon_2\cos\phi)^{2\alpha+1}}d\phi, \tag{27}$$

where $\alpha$ and $\varepsilon_0$ are power and ellipticity parameters respectively, which take different values, obtained by Glauber model for 5.02 GeV Pb-Pb collisions, depending on the centrality [19]. The scaling factor $\kappa_2$ between the elliptic flow and the initial eccentricity, $v_2 = \kappa_2\varepsilon_2$, is chosen to imitate the centrality dependence of the elliptic flow $v_2$ measured in Ref. [16]. For each event of a given $v_2$ a simple distribution $1 + 2v_2\cos(2\phi)$ is used to generate particle azimuthal angle. The integral in Eq. (27) can be carried out analytically to give [20]:

$$\frac{dN}{d\varepsilon_2} = 2\alpha(1-\varepsilon_0^2)^{\alpha+1/2}\varepsilon_2 \frac{(1-\varepsilon_2^2)^{\alpha-1}}{(1-\varepsilon_0\varepsilon_2)^{2\alpha+1}} \, {}_2F_1\left(\frac{1}{2}, 2\alpha+1; 1; \frac{2\varepsilon_0\varepsilon_2}{\varepsilon_0\varepsilon_2 - 1}\right). \tag{28}$$

However, the ROOT version [21] of the hypergeometric function in Eq. (28) is not defined everywhere in the interval (0, 1) of $\varepsilon_2$. In Fig. 2 is shown the parametric area of definition of the ROOT version of the hypergeometric function.

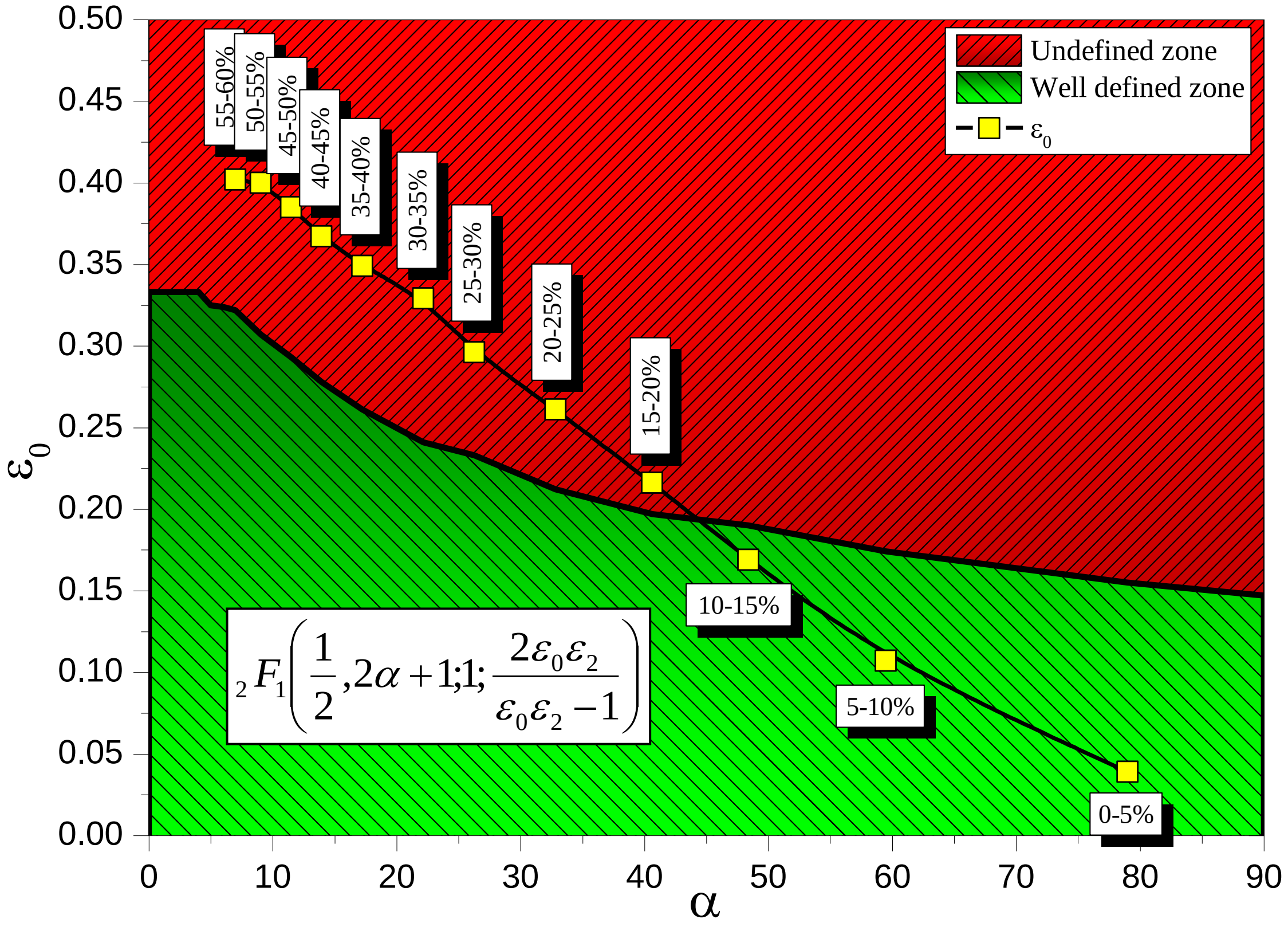

FIG. 2 Parameter values ($\alpha$, $\varepsilon_0$) of the elliptic power distribution for 5.02 GeV Pb-Pb collisions calculated in the Glauber model for different centralities, and the area of definition of the ROOT version of the hypergeometric function ${}_2F_1\left(\frac{1}{2}, 2\alpha+1; 1; \frac{2\varepsilon_0\varepsilon_2}{\varepsilon_0\varepsilon_2 - 1}\right)$.

It can be seen that only the centrality range 0-15% of the 5.02 GeV Pb-Pb collisions might be well simulated by this hypergeometric function. We solved this inconvenience by applying the Pfaff transformation:

$${}_2F_1(a,b;c;z) = (1-z)^{-b} \, {}_2F_1\left(c-a, b; c; \frac{z}{z-1}\right), \tag{29}$$

which gives the following eccentricity distribution, well defined for all parameter values:

$$\frac{dN}{d\varepsilon_2} = 2\alpha(1-\varepsilon_0^2)^{\alpha+1/2}\varepsilon_2 \frac{(1-\varepsilon_2^2)^{\alpha-1}}{(1+\varepsilon_0\varepsilon_2)^{2\alpha+1}} \; {}_2F_1\left(\frac{1}{2}, 2\alpha+1; 1; \frac{2\varepsilon_0\varepsilon_2}{1+\varepsilon_0\varepsilon_2}\right). \quad (30)$$

For each centrality, about 1.5 x $10^7$ events have been simulated using the above described toy model. Figure 3 shows $v_2\{2k\}$ values ($k$ = 1,…,7) calculated based on the obtained expressions of the corresponding $\langle 2m \rangle$ correlations as a function of centrality. A gap between the $v_2\{2\}$ and higher-order cumulant values $v_2\{2k\}$ ($k$ = 2,…,7) is present. It is due to flow fluctuations that relates higher-order cumulants based $v_2\{2k\}$ and the variance $\sigma_v$ of the $v_2$ distribution: $v_2\{2\}^2 \approx v_2\{2k\}^2 + 2\sigma^2_v$, for $k$ > 1 [15]. As it has been seen from the experimental data [16], flow fluctuations become larger going to peripheral collisions. The elliptic flow $v_2$ values are reproduced using the expressions for the $\langle 2m \rangle$ developed in this paper.

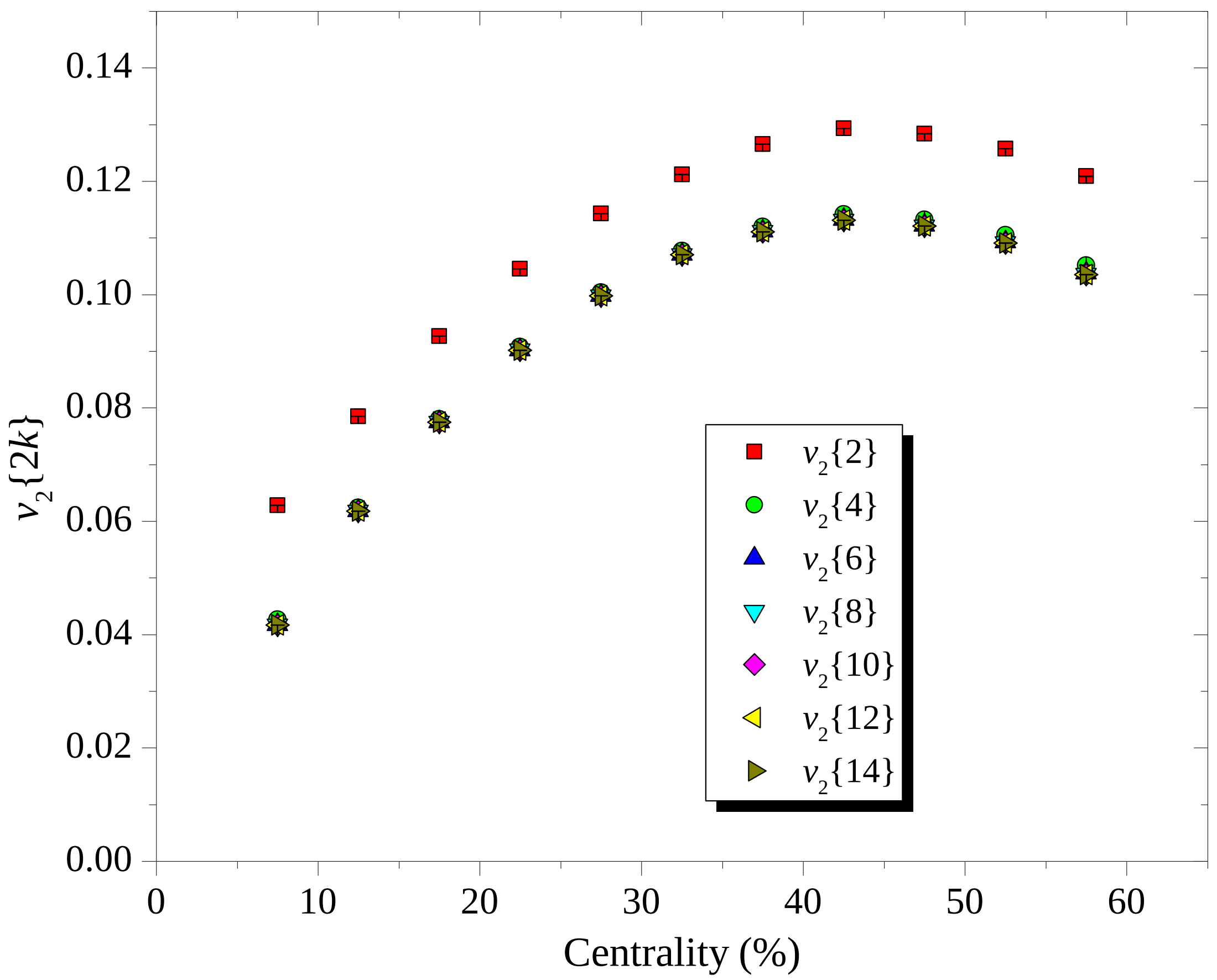

FIG. 3 The $v_2\{2k\}$ values ($k$ = 1,…,7) as a function of centrality calculated from the data simulated by the elliptic power distribution toy model. The statistical uncertainties are negligible compared to the marker size.

As the fluctuations in the initial state are not Gaussian, the $v_2\{2k\}$ values for $k$ > 1 will not be the same. This will produce a splitting between different $v_2\{2k\}$ values and they will be ordered as $v_2\{2k\}$ > $v_2\{2(k + 1)\}$ for any $k$ > 1. In Fig. 3 the ordering and the fine splitting between the $v_2\{2k\}$ values ($k$ = 2,…,7) are not well visible. In order to make the splitting between the cumulants of different orders explicitly visible, we show in Fig. 4 the relative differences $(v_2\{2k\} - v_2\{14\})/v_2\{14\}$ ($k$ = 1,…,6) as a function of centrality. The points in Fig. 4 depicted with different symbols for different cumulant orders are calculated from the simulated data. On the other hand, the corresponding input values, obtained directly applying the elliptic power distributions, are represented by spline interpolation lines. In Fig. 4 the ordering is clearly seen, as well as the fine splitting between the cumulants of different orders. The relative difference between the cumulants decreases by about one order of magnitude for each increment of the order $k$. An excellent agreement between the lines and the symbols prove correctness of the obtained expressions for the 2$m$-particle azimuthal correlations $\langle 2m \rangle$.

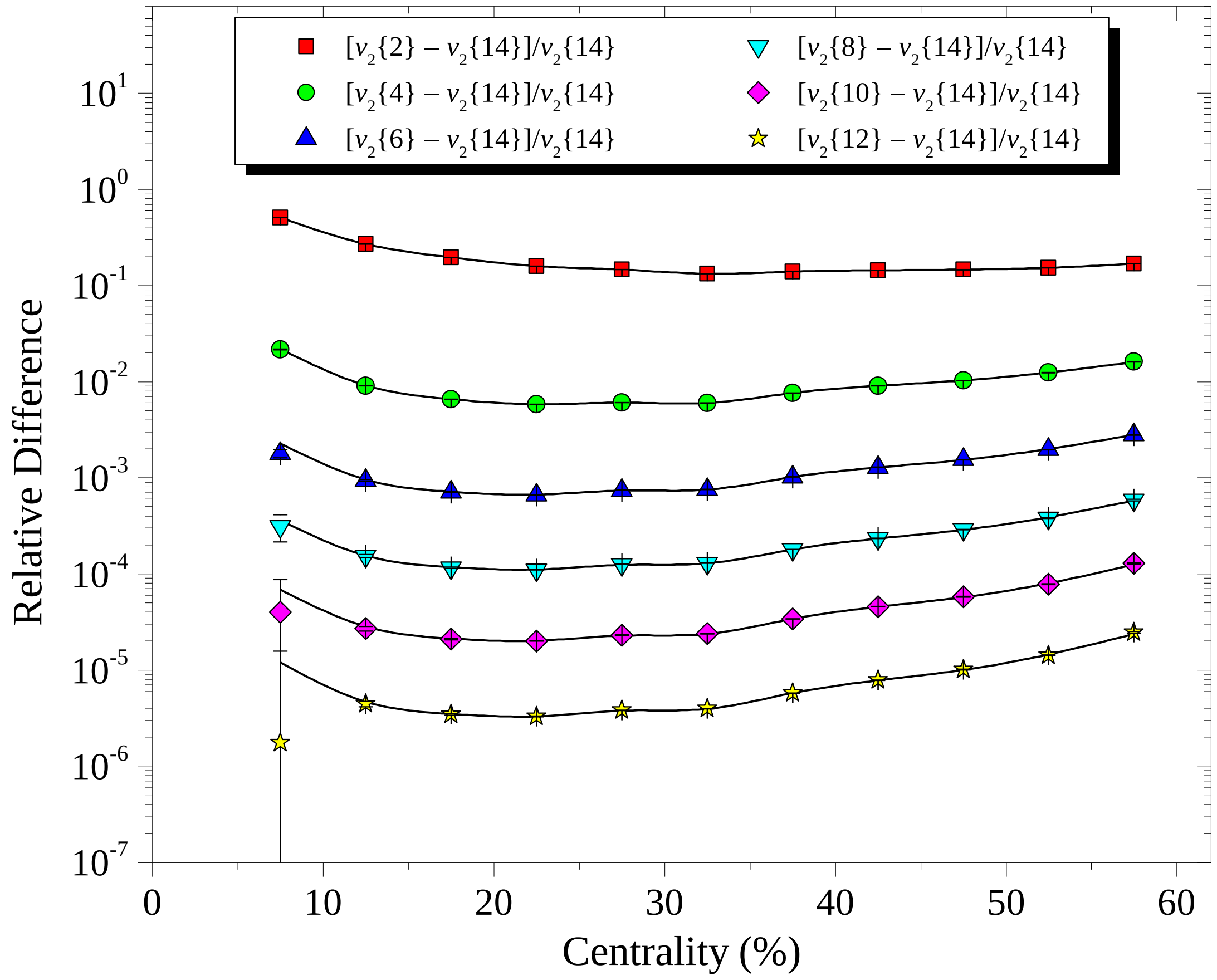

FIG. 4 The relative differences $(v_2\{2k\} - v_2\{14\})/v_2\{14\}$ ($k$ = 1,…,6) as a function of centrality. The input values, obtained directly applying the elliptic power distributions, are represented by spline interpolation lines.

## Conclusions

In this paper we present a method, based on mathematical induction, that allows to calculate the high-order $v_2\{2k\}$ values from the Q-cumulants in a relatively straightforward way. The analytical expressions for high 2$m$-particle correlations, $\langle 10 \rangle$, $\langle 12 \rangle$, and $\langle 14 \rangle$, are obtained for the first time and given here in the form of table values. The validity of the proposed method is confirmed by the elliptic flow simulation with a toy model that exploits elliptic power distribution. We transformed the hypergeometric function, involved in the elliptic power distribution, by applying Pfaff transformation, and enabled its calculation in the ROOT program for all parameter values of the $v_2$ distribution. The ability to calculate high-order $v_2\{2k\}$ values allows the possibility to study the fine details of the $v_2$ distribution by extracting its skewness and higher central moments. Theoreticians can tune initial states in their hydrodynamics models such to reconstruct measured central moments. This can place stringent constraints on the initial geometry used in hydrodynamic calculations of the collective dynamics of the QGP in high energy nuclear collisions.

## Acknowledgments

The authors acknowledge the support from Ministry of Education Science and Technological Development, Republic of Serbia, National Natural Science Foundation of China (Grant No. 12035006, 12075085, 12147219) and the U.S. Department of Energy (Grant No. de-sc0012910).

## Appendix

This Appendix provides Tables listing, for different orders of the Q-cumulants, the coefficients, integer functions and combinations of the basis vectors needed for the construction of expressions for the $2m$-particle azimuthal correlations $\langle 2m \rangle$.

**Table 3** Coefficients, integer functions and basis vectors for calculation of the six-particle azimuthal correlations.

| $i$ | $x_i$ | $f_i^{(m=3,l)}$ | **Basis vectors** | $l$ |
|---|---|---|---|---|
| 1 | -6 | $M(M-4)(M-5)$ | $1$ | 0 |
| 2 | 18 | $(M-2)(M-5)$ | $Q_n \mid Q_n^*$ | 1 |
| 3 | -9 | $(M-4)$ | $Q_nQ_n \mid Q_n^*Q_n^*$ | 2 |
| 4 | 18 | $(M-4)$ | $Q_{2n} \mid Q_n^*Q_n^*$ | 2 |
| 5 | -9 | $(M-4)$ | $Q_{2n} \mid Q_{2n}^*$ | 2 |
| 6 | 1 | 1 | $Q_nQ_nQ_n \mid Q_n^*Q_n^*Q_n^*$ | 3 |
| 7 | -6 | 1 | $Q_{2n}\ Q_n \mid Q_n^*Q_n^*Q_n^*$ | 3 |
| 8 | 9 | 1 | $Q_{2n}\ Q_n \mid Q_n^*\ Q_{2n}^*$ | 3 |
| 9 | 4 | 1 | $Q_{3n} \mid Q_n^*Q_n^*Q_n^*$ | 3 |
| 10 | -12 | 1 | $Q_{3n} \mid Q_n^*\ Q_{2n}^*$ | 3 |
| 11 | 4 | 1 | $Q_{3n} \mid Q_{3n}^*$ | 3 |

**Table 4** Coefficients, integer functions and basis vectors for calculation of the eight-particle azimuthal correlations.

| $i$ | $x_i$ | $f_i^{(m=4,l)}$ | **Basis vectors** | $l$ |
|---|---|---|---|---|
| 1 | 24 | $M(M-5)(M-6)(M-7)$ | $1$ | 0 |
| 2 | -96 | $(M-2)(M-6)(M-7)$ | $Q_n \mid Q_n^*$ | 1 |
| 3 | 72 | $(M-4)(M-7)$ | $Q_nQ_n \mid Q_n^*Q_n^*$ | 2 |
| 4 | -144 | $(M-4)(M-7)$ | $Q_{2n} \mid Q_n^*Q_n^*$ | 2 |
| 5 | 72 | $(M-4)(M-7)$ | $Q_{2n} \mid Q_{2n}^*$ | 2 |
| 6 | -16 | $(M-6)$ | $Q_nQ_nQ_n \mid Q_n^*Q_n^*Q_n^*$ | 3 |
| 7 | 96 | $(M-6)$ | $Q_{2n}\ Q_n \mid Q_n^*Q_n^*Q_n^*$ | 3 |
| 8 | -144 | $(M-6)$ | $Q_{2n}\ Q_n \mid Q_n^*\ Q_{2n}^*$ | 3 |
| 9 | -64 | $(M-6)$ | $Q_{3n} \mid Q_n^*Q_n^*Q_n^*$ | 3 |
| 10 | 192 | $(M-6)$ | $Q_{3n} \mid Q_n^*\ Q_{2n}^*$ | 3 |
| 11 | -64 | $(M-6)$ | $Q_{3n} \mid Q_{3n}^*$ | 3 |
| 12 | 1 | 1 | $Q_nQ_nQ_nQ_n \mid Q_n^*Q_n^*Q_n^*Q_n^*$ | 4 |
| 13 | -12 | 1 | $Q_{2n}\ Q_nQ_n \mid Q_n^*Q_n^*Q_n^*Q_n^*$ | 4 |
| 14 | 36 | 1 | $Q_{2n}\ Q_nQ_n \mid Q_n^*Q_n^*\ Q_{2n}^*$ | 4 |
| 15 | 6 | 1 | $Q_{2n}\ Q_{2n} \mid Q_n^*Q_n^*Q_n^*Q_n^*$ | 4 |
| 16 | -36 | 1 | $Q_{2n}\ Q_{2n} \mid Q_n^*Q_n^*\ Q_{2n}^*$ | 4 |
| 17 | 9 | 1 | $Q_{2n} Q_{2n} \mid Q^*_{2n} Q^*_{2n}$ | 4 |
| 18 | 16 | 1 | $Q_{3n} Q_n \mid Q^*_n Q^*_n Q^*_n Q^*_n$ | 4 |
| 19 | -96 | 1 | $Q_{3n} Q_n \mid Q^*_n Q^*_n Q^*_{2n}$ | 4 |
| 20 | 48 | 1 | $Q_{3n} Q_n \mid Q^*_{2n} Q^*_{2n}$ | 4 |
| 21 | 64 | 1 | $Q_{3n} Q_n \mid Q^*_n Q^*_{3n}$ | 4 |
| 22 | -12 | 1 | $Q_{4n} \mid Q^*_n Q^*_n Q^*_n Q^*_n$ | 4 |
| 23 | 72 | 1 | $Q_{4n} \mid Q^*_n Q^*_n Q^*_{2n}$ | 4 |
| 24 | -36 | 1 | $Q_{4n} \mid Q^*_{2n} Q^*_{2n}$ | 4 |
| 25 | -96 | 1 | $Q_{4n} \mid Q^*_n Q^*_{3n}$ | 4 |
| 26 | 36 | 1 | $Q_{4n} \mid Q^*_{4n}$ | 4 |

**Table 5** Coefficients, integer functions and basis vectors for calculation of the ten-particle azimuthal correlations.

| $i$ | $x_i$ | $f_i^{(m=5,l)}$ | **Basis vectors** | $l$ |
|---|---|---|---|---|
| 1 | -120 | $M(M-6)(M-7)(M-8)(M-9)$ | 1 | 0 |
| 2 | 600 | $(M-2)(M-7)(M-8)(M-9)$ | $Q_n \mid Q^*_n$ | 1 |
| 3 | -600 | $(M-4)(M-8)(M-9)$ | $Q_n Q_n \mid Q^*_n Q^*_n$ | 2 |
| 4 | 1200 | $(M-4)(M-8)(M-9)$ | $Q_{2n} \mid Q^*_n Q^*_n$ | 2 |
| 5 | -600 | $(M-4)(M-8)(M-9)$ | $Q_{2n} \mid Q^*_{2n}$ | 2 |
| 6 | 200 | $(M-6)(M-9)$ | $Q_n Q_n Q_n \mid Q^*_n Q^*_n Q^*_n$ | 3 |
| 7 | -1200 | $(M-6)(M-9)$ | $Q_{2n} Q_n \mid Q^*_n Q^*_n Q^*_n$ | 3 |
| 8 | 1800 | $(M-6)(M-9)$ | $Q_{2n} Q_n \mid Q^*_n Q^*_{2n}$ | 3 |
| 9 | 800 | $(M-6)(M-9)$ | $Q_{3n} \mid Q^*_n Q^*_n Q^*_n$ | 3 |
| 10 | -2400 | $(M-6)(M-9)$ | $Q_{3n} \mid Q^*_n Q^*_{2n}$ | 3 |
| 11 | 800 | $(M-6)(M-9)$ | $Q_{3n} \mid Q^*_{3n}$ | 3 |
| 12 | -25 | $(M-8)$ | $Q_n Q_n Q_n Q_n \mid Q^*_n Q^*_n Q^*_n Q^*_n$ | 4 |
| 13 | 300 | $(M-8)$ | $Q_{2n} Q_n Q_n \mid Q^*_n Q^*_n Q^*_n Q^*_n$ | 4 |
| 14 | -900 | $(M-8)$ | $Q_{2n} Q_n Q_n \mid Q^*_n Q^*_n Q^*_{2n}$ | 4 |
| 15 | -150 | $(M-8)$ | $Q_{2n} Q_{2n} \mid Q^*_n Q^*_n Q^*_n Q^*_n$ | 4 |
| 16 | 900 | $(M-8)$ | $Q_{2n} Q_{2n} \mid Q^*_n Q^*_n Q^*_{2n}$ | 4 |
| 17 | -225 | $(M-8)$ | $Q_{2n} Q_{2n} \mid Q^*_{2n} Q^*_{2n}$ | 4 |
| 18 | -400 | $(M-8)$ | $Q_{3n} Q_n \mid Q^*_n Q^*_n Q^*_n Q^*_n$ | 4 |
| 19 | 2400 | $(M-8)$ | $Q_{3n} Q_n \mid Q^*_n Q^*_n Q^*_{2n}$ | 4 |
| 20 | -1200 | $(M-8)$ | $Q_{3n} Q_n \mid Q^*_{2n} Q^*_{2n}$ | 4 |
| 21 | -1600 | $(M-8)$ | $Q_{3n} Q_n \mid Q^*_n Q^*_{3n}$ | 4 |
| 22 | 300 | $(M-8)$ | $Q_{4n} \mid Q_n^* Q_n^* Q_n^* Q_n^*$ | 4 |
| 23 | -1800 | $(M-8)$ | $Q_{4n} \mid Q_n^* Q_n^* \; Q_{2n}^*$ | 4 |
| 24 | 900 | $(M-8)$ | $Q_{4n} \mid Q_{2n}^* \; Q_{2n}^*$ | 4 |
| 25 | 2400 | $(M-8)$ | $Q_{4n} \mid Q_n^* \; Q_{3n}^*$ | 4 |
| 26 | -900 | $(M-8)$ | $Q_{4n} \mid Q_{4n}^*$ | 4 |
| 27 | 1 | 1 | $Q_n Q_n Q_n Q_n Q_n \mid Q_n^* Q_n^* Q_n^* Q_n^* Q_n^*$ | 5 |
| 28 | -20 | 1 | $Q_{2n} \; Q_n Q_n Q_n \mid Q_n^* Q_n^* Q_n^* Q_n^* Q_n^*$ | 5 |
| 29 | 100 | 1 | $Q_{2n} \; Q_n Q_n Q_n \mid Q_n^* Q_n^* Q_n^* \; Q_{2n}^*$ | 5 |
| 30 | 30 | 1 | $Q_{2n} \; Q_{2n} \; Q_n \mid Q_n^* Q_n^* Q_n^* Q_n^* Q_n^*$ | 5 |
| 31 | -300 | 1 | $Q_{2n} \; Q_{2n} \; Q_n \mid Q_n^* Q_n^* Q_n^* \; Q_{2n}^*$ | 5 |
| 32 | 225 | 1 | $Q_{2n} \; Q_{2n} \; Q_n \mid Q_n^* \; Q_{2n}^* \; Q_{2n}^*$ | 5 |
| 33 | 40 | 1 | $Q_{3n} \; Q_n Q_n \mid Q_n^* Q_n^* Q_n^* Q_n^* Q_n^*$ | 5 |
| 34 | -400 | 1 | $Q_{3n} \; Q_n Q_n \mid Q_n^* Q_n^* Q_n^* \; Q_{2n}^*$ | 5 |
| 35 | 600 | 1 | $Q_{3n} \; Q_n Q_n \mid Q_n^* \; Q_{2n}^* \; Q_{2n}^*$ | 5 |
| 36 | 400 | 1 | $Q_{3n} \; Q_n Q_n \mid Q_n^* Q_n^* \; Q_{3n}^*$ | 5 |
| 37 | -40 | 1 | $Q_{3n} \; Q_{2n} \mid Q_n^* Q_n^* Q_n^* Q_n^* Q_n^*$ | 5 |
| 38 | 400 | 1 | $Q_{3n} \; Q_{2n} \mid Q_n^* Q_n^* Q_n^* \; Q_{2n}^*$ | 5 |
| 39 | -600 | 1 | $Q_{3n} \; Q_{2n} \mid Q_n^* \; Q_{2n}^* \; Q_{2n}^*$ | 5 |
| 40 | -800 | 1 | $Q_{3n} \; Q_{2n} \mid Q_n^* Q_n^* \; Q_{3n}^*$ | 5 |
| 41 | 400 | 1 | $Q_{3n} \; Q_{2n} \mid Q_{2n}^* \; Q_{3n}^*$ | 5 |
| 42 | -60 | 1 | $Q_{4n} \; Q_n \mid Q_n^* Q_n^* Q_n^* Q_n^* Q_n^*$ | 5 |
| 43 | 600 | 1 | $Q_{4n} \; Q_n \mid Q_n^* Q_n^* Q_n^* \; Q_{2n}^*$ | 5 |
| 44 | -900 | 1 | $Q_{4n} \; Q_n \mid Q_n^* \; Q_{2n}^* \; Q_{2n}^*$ | 5 |
| 45 | -1200 | 1 | $Q_{4n} \; Q_n \mid Q_n^* Q_n^* \; Q_{3n}^*$ | 5 |
| 46 | 1200 | 1 | $Q_{4n} \; Q_n \mid Q_{2n}^* \; Q_{3n}^*$ | 5 |
| 47 | 900 | 1 | $Q_{4n} \; Q_n \mid Q_n^* \; Q_{4n}^*$ | 5 |
| 48 | 48 | 1 | $Q_{5n} \mid Q_n^* Q_n^* Q_n^* Q_n^* Q_n^*$ | 5 |
| 49 | -480 | 1 | $Q_{5n} \mid Q_n^* Q_n^* Q_n^* \; Q_{2n}^*$ | 5 |
| 50 | 720 | 1 | $Q_{5n} \mid Q_n^* \; Q_{2n}^* \; Q_{2n}^*$ | 5 |
| 51 | 960 | 1 | $Q_{5n} \mid Q_n^* Q_n^* \; Q_{3n}^*$ | 5 |
| 52 | -960 | 1 | $Q_{5n} \mid Q_{2n}^* \; Q_{3n}^*$ | 5 |
| 53 | -1440 | 1 | $Q_{5n} \mid Q_n^* \; Q_{4n}^*$ | 5 |
| 54 | 576 | 1 | $Q_{5n} \mid Q_{5n}^*$ | 5 |

**Table 6** Coefficients, integer functions and basis vectors for calculation of the twelve-particle azimuthal correlations.

| $i$ | $x_i$ | $f_i^{(m=6,l)}$ | **Basis vectors** | $l$ |
|---|---|---|---|---|
| 1 | 720 | $M(M-7)(M-8)(M-9)(M-10)(M-11)$ | $1$ | 0 |
| 2 | -4320 | $(M-2)(M-8)(M-9)(M-10)(M-11)$ | $Q_n \mid Q_n^*$ | 1 |
| 3 | 5400 | $(M-4)(M-9)(M-10)(M-11)$ | $Q_nQ_n \mid Q_n^*Q_n^*$ | 2 |
| 4 | -10800 | $(M-4)(M-9)(M-10)(M-11)$ | $Q_{2n} \mid Q_n^*Q_n^*$ | 2 |
| 5 | 5400 | $(M-4)(M-9)(M-10)(M-11)$ | $Q_{2n} \mid Q_{2n}^*$ | 2 |
| 6 | -2400 | $(M-6)(M-10)(M-11)$ | $Q_nQ_nQ_n \mid Q_n^*Q_n^*Q_n^*$ | 3 |
| 7 | 14400 | $(M-6)(M-10)(M-11)$ | $Q_{2n}Q_n \mid Q_n^*Q_n^*Q_n^*$ | 3 |
| 8 | -21600 | $(M-6)(M-10)(M-11)$ | $Q_{2n}Q_n \mid Q_n^*Q_{2n}^*$ | 3 |
| 9 | -9600 | $(M-6)(M-10)(M-11)$ | $Q_{3n} \mid Q_n^*Q_n^*Q_n^*$ | 3 |
| 10 | 28800 | $(M-6)(M-10)(M-11)$ | $Q_{3n} \mid Q_n^*Q_{2n}^*$ | 3 |
| 11 | -9600 | $(M-6)(M-10)(M-11)$ | $Q_{3n} \mid Q_{3n}^*$ | 3 |
| 12 | 450 | $(M-8)(M-11)$ | $Q_nQ_nQ_nQ_n \mid Q_n^*Q_n^*Q_n^*Q_n^*$ | 4 |
| 13 | -5400 | $(M-8)(M-11)$ | $Q_{2n}Q_nQ_n \mid Q_n^*Q_n^*Q_n^*Q_n^*$ | 4 |
| 14 | 16200 | $(M-8)(M-11)$ | $Q_{2n}Q_nQ_n \mid Q_n^*Q_n^*Q_{2n}^*$ | 4 |
| 15 | 2700 | $(M-8)(M-11)$ | $Q_{2n}Q_{2n} \mid Q_n^*Q_n^*Q_n^*Q_n^*$ | 4 |
| 16 | -16200 | $(M-8)(M-11)$ | $Q_{2n}Q_{2n} \mid Q_n^*Q_n^*Q_{2n}^*$ | 4 |
| 17 | 4050 | $(M-8)(M-11)$ | $Q_{2n}Q_{2n} \mid Q_{2n}^*Q_{2n}^*$ | 4 |
| 18 | 7200 | $(M-8)(M-11)$ | $Q_{3n}Q_n \mid Q_n^*Q_n^*Q_n^*Q_n^*$ | 4 |
| 19 | -43200 | $(M-8)(M-11)$ | $Q_{3n}Q_n \mid Q_n^*Q_n^*Q_{2n}^*$ | 4 |
| 20 | 21600 | $(M-8)(M-11)$ | $Q_{3n}Q_n \mid Q_{2n}^*Q_{2n}^*$ | 4 |
| 21 | 28800 | $(M-8)(M-11)$ | $Q_{3n}Q_n \mid Q_n^*Q_{3n}^*$ | 4 |
| 22 | -5400 | $(M-8)(M-11)$ | $Q_{4n} \mid Q_n^*Q_n^*Q_n^*Q_n^*$ | 4 |
| 23 | 32400 | $(M-8)(M-11)$ | $Q_{4n} \mid Q_n^*Q_n^*Q_{2n}^*$ | 4 |
| 24 | -16200 | $(M-8)(M-11)$ | $Q_{4n} \mid Q_{2n}^*Q_{2n}^*$ | 4 |
| 25 | -43200 | $(M-8)(M-11)$ | $Q_{4n} \mid Q_n^*Q_{3n}^*$ | 4 |
| 26 | 16200 | $(M-8)(M-11)$ | $Q_{4n} \mid Q_{4n}^*$ | 4 |
| 27 | -36 | $(M-10)$ | $Q_nQ_nQ_nQ_nQ_n \mid Q_n^*Q_n^*Q_n^*Q_n^*Q_n^*$ | 5 |
| 28 | 720 | $(M-10)$ | $Q_{2n}Q_nQ_nQ_n \mid Q_n^*Q_n^*Q_n^*Q_n^*Q_n^*$ | 5 |
| 29 | -3600 | $(M-10)$ | $Q_{2n}Q_nQ_nQ_n \mid Q_n^*Q_n^*Q_n^*Q_{2n}^*$ | 5 |
| 30 | -1080 | $(M-10)$ | $Q_{2n}Q_{2n}Q_n \mid Q_n^*Q_n^*Q_n^*Q_n^*Q_n^*$ | 5 |
| 31 | 10800 | $(M-10)$ | $Q_{2n}Q_{2n}Q_n \mid Q_n^*Q_n^*Q_n^*Q_{2n}^*$ | 5 |
| 32 | -8100 | $(M-10)$ | $Q_{2n}Q_{2n}Q_n \mid Q_n^*Q_{2n}^*Q_{2n}^*$ | 5 |
| 33 | -1440 | $(M-10)$ | $Q_{3n} \quad Q_n Q_n \mid Q_n^* Q_n^* Q_n^* Q_n^* Q_n^*$ | 5 |
| 34 | 14400 | $(M-10)$ | $Q_{3n} \quad Q_n Q_n \mid Q_n^* Q_n^* Q_n^* \quad Q_{2n}^*$ | 5 |
| 35 | -21600 | $(M-10)$ | $Q_{3n} \quad Q_n Q_n \mid Q_n^* \quad Q_{2n}^* \quad Q_{2n}^*$ | 5 |
| 36 | -14400 | $(M-10)$ | $Q_{3n} \quad Q_n Q_n \mid Q_n^* Q_n^* \quad Q_{3n}^*$ | 5 |
| 37 | 1440 | $(M-10)$ | $Q_{3n} \quad Q_{2n} \mid Q_n^* Q_n^* Q_n^* Q_n^* Q_n^*$ | 5 |
| 38 | -14400 | $(M-10)$ | $Q_{3n} \quad Q_{2n} \mid Q_n^* Q_n^* Q_n^* \quad Q_{2n}^*$ | 5 |
| 39 | 21600 | $(M-10)$ | $Q_{3n} \quad Q_{2n} \mid Q_n^* \quad Q_{2n}^* \quad Q_{2n}^*$ | 5 |
| 40 | 28800 | $(M-10)$ | $Q_{3n} \quad Q_{2n} \mid Q_n^* Q_n^* \quad Q_{3n}^*$ | 5 |
| 41 | -14400 | $(M-10)$ | $Q_{3n} \quad Q_{2n} \mid \quad Q_{2n}^* \quad Q_{3n}^*$ | 5 |
| 42 | 2160 | $(M-10)$ | $Q_{4n} \quad Q_n \mid Q_n^* Q_n^* Q_n^* Q_n^* Q_n^*$ | 5 |
| 43 | -21600 | $(M-10)$ | $Q_{4n} \quad Q_n \mid Q_n^* Q_n^* Q_n^* \quad Q_{2n}^*$ | 5 |
| 44 | 32400 | $(M-10)$ | $Q_{4n} \quad Q_n \mid Q_n^* \quad Q_{2n}^* \quad Q_{2n}^*$ | 5 |
| 45 | 43200 | $(M-10)$ | $Q_{4n} \quad Q_n \mid Q_n^* Q_n^* \quad Q_{3n}^*$ | 5 |
| 46 | -43200 | $(M-10)$ | $Q_{4n} \quad Q_n \mid \quad Q_{2n}^* \quad Q_{3n}^*$ | 5 |
| 47 | -32400 | $(M-10)$ | $Q_{4n} \quad Q_n \mid Q_n^* \quad Q_{4n}^*$ | 5 |
| 48 | -1728 | $(M-10)$ | $Q_{5n} \quad \mid Q_n^* Q_n^* Q_n^* Q_n^* Q_n^*$ | 5 |
| 49 | 17280 | $(M-10)$ | $Q_{5n} \quad \mid Q_n^* Q_n^* Q_n^* \quad Q_{2n}^*$ | 5 |
| 50 | -25920 | $(M-10)$ | $Q_{5n} \quad \mid Q_n^* \quad Q_{2n}^* \quad Q_{2n}^*$ | 5 |
| 51 | -34560 | $(M-10)$ | $Q_{5n} \quad \mid Q_n^* Q_n^* \quad Q_{3n}^*$ | 5 |
| 52 | 34560 | $(M-10)$ | $Q_{5n} \quad \mid \quad Q_{2n}^* \quad Q_{3n}^*$ | 5 |
| 53 | 51840 | $(M-10)$ | $Q_{5n} \quad \mid Q_n^* \quad Q_{4n}^*$ | 5 |
| 54 | -20736 | $(M-10)$ | $Q_{5n} \quad \mid \quad Q_{5n}^*$ | 5 |
| 55 | 1 | 1 | $Q_n Q_n Q_n Q_n Q_n Q_n \mid Q_n^* Q_n^* Q_n^* Q_n^* Q_n^* Q_n^*$ | 6 |
| 56 | -30 | 1 | $Q_{2n} \quad Q_n Q_n Q_n Q_n \mid Q_n^* Q_n^* Q_n^* Q_n^* Q_n^* Q_n^*$ | 6 |
| 57 | 225 | 1 | $Q_{2n} \quad Q_n Q_n Q_n Q_n \mid Q_n^* Q_n^* Q_n^* Q_n^* \quad Q_{2n}^*$ | 6 |
| 58 | 90 | 1 | $Q_{2n} \quad Q_{2n} \quad Q_n Q_n \mid Q_n^* Q_n^* Q_n^* Q_n^* Q_n^* Q_n^*$ | 6 |
| 59 | -1350 | 1 | $Q_{2n} \quad Q_{2n} \quad Q_n Q_n \mid Q_n^* Q_n^* Q_n^* Q_n^* \quad Q_{2n}^*$ | 6 |
| 60 | 2025 | 1 | $Q_{2n} \quad Q_{2n} \quad Q_n Q_n \mid Q_n^* Q_n^* \quad Q_{2n}^* \quad Q_{2n}^*$ | 6 |
| 61 | -30 | 1 | $Q_{2n} \quad Q_{2n} \quad Q_{2n} \mid Q_n^* Q_n^* Q_n^* Q_n^* Q_n^* Q_n^*$ | 6 |
| 62 | 450 | 1 | $Q_{2n} \quad Q_{2n} \quad Q_{2n} \mid Q_n^* Q_n^* Q_n^* Q_n^* \quad Q_{2n}^*$ | 6 |
| 63 | -1350 | 1 | $Q_{2n} \quad Q_{2n} \quad Q_{2n} \mid Q_n^* Q_n^* \quad Q_{2n}^* \quad Q_{2n}^*$ | 6 |
| 64 | 225 | 1 | $Q_{2n} \quad Q_{2n} \quad Q_{2n} \mid Q_{2n}^* \quad Q_{2n}^* \quad Q_{2n}^*$ | 6 |
| 65 | 80 | 1 | $Q_{3n} \quad Q_n Q_n Q_n \mid Q_n^* Q_n^* Q_n^* Q_n^* Q_n^* Q_n^*$ | 6 |
| 66 | -1200 | 1 | $Q_{3n} \quad Q_n Q_n Q_n \mid Q_n^* Q_n^* Q_n^* Q_n^* \quad Q_{2n}^*$ | 6 |
| 67 | 3600 | 1 | $Q_{3n} \quad Q_n Q_n Q_n \mid Q_n^* Q_n^* \quad Q_{2n}^* \quad Q_{2n}^*$ | 6 |
| 68 | -1200 | 1 | $Q_{3n} \quad Q_n Q_n Q_n \mid Q_{2n}^* \quad Q_{2n}^* \quad Q_{2n}^*$ | 6 |
| 69 | 1600 | 1 | $Q_{3n} \quad Q_n Q_n Q_n \mid Q_n^* Q_n^* Q_n^* \quad Q_{3n}^*$ | 6 |
| 70 | -240 | 1 | $Q_{3n} \quad Q_{2n} \quad Q_n \mid Q_n^* Q_n^* Q_n^* Q_n^* Q_n^* Q_n^*$ | 6 |
| 71 | 3600 | 1 | $Q_{3n} \quad Q_{2n} \quad Q_n \mid Q_n^* Q_n^* Q_n^* Q_n^* \quad Q_{2n}^*$ | 6 |
| 72 | -10800 | 1 | $Q_{3n} \quad Q_{2n} \quad Q_n \mid Q_n^* Q_n^* \quad Q_{2n}^* \quad Q_{2n}^*$ | 6 |
| 73 | 3600 | 1 | $Q_{3n} \quad Q_{2n} \quad Q_n \mid \quad Q_{2n}^* \quad Q_{2n}^* \quad Q_{2n}^*$ | 6 |
| 74 | -9600 | 1 | $Q_{3n} \quad Q_{2n} \quad Q_n \mid Q_n^* Q_n^* Q_n^* \quad Q_{3n}^*$ | 6 |
| 75 | 14400 | 1 | $Q_{3n} \quad Q_{2n} \quad Q_n \mid Q_n^* \quad Q_{2n}^* \quad Q_{3n}^*$ | 6 |
| 76 | 80 | 1 | $Q_{3n} \quad Q_{3n} \quad \mid Q_n^* Q_n^* Q_n^* Q_n^* Q_n^* Q_n^*$ | 6 |
| 77 | -1200 | 1 | $Q_{3n} \quad Q_{3n} \quad \mid Q_n^* Q_n^* Q_n^* Q_n^* \quad Q_{2n}^*$ | 6 |
| 78 | 3600 | 1 | $Q_{3n} \quad Q_{3n} \quad \mid Q_n^* Q_n^* \quad Q_{2n}^* \quad Q_{2n}^*$ | 6 |
| 79 | -1200 | 1 | $Q_{3n} \quad Q_{3n} \quad \mid \quad Q_{2n}^* \quad Q_{2n}^* \quad Q_{2n}^*$ | 6 |
| 80 | 3200 | 1 | $Q_{3n} \quad Q_{3n} \quad \mid Q_n^* Q_n^* Q_n^* \quad Q_{3n}^*$ | 6 |
| 81 | -9600 | 1 | $Q_{3n} \quad Q_{3n} \quad \mid Q_n^* \quad Q_{2n}^* \quad Q_{3n}^*$ | 6 |
| 82 | 1600 | 1 | $Q_{3n} \quad Q_{3n} \quad \mid \quad Q_{3n}^* \quad Q_{3n}^*$ | 6 |
| 83 | -180 | 1 | $Q_{4n} \quad Q_n Q_n \mid Q_n^* Q_n^* Q_n^* Q_n^* Q_n^* Q_n^*$ | 6 |
| 84 | 2700 | 1 | $Q_{4n} \quad Q_n Q_n \mid Q_n^* Q_n^* Q_n^* Q_n^* \quad Q_{2n}^*$ | 6 |
| 85 | -8100 | 1 | $Q_{4n} \quad Q_n Q_n \mid Q_n^* Q_n^* \quad Q_{2n}^* \quad Q_{2n}^*$ | 6 |
| 86 | 2700 | 1 | $Q_{4n} \quad Q_n Q_n \mid \quad Q_{2n}^* \quad Q_{2n}^* \quad Q_{2n}^*$ | 6 |
| 87 | -7200 | 1 | $Q_{4n} \quad Q_n Q_n \mid Q_n^* Q_n^* Q_n^* \quad Q_{3n}^*$ | 6 |
| 88 | 21600 | 1 | $Q_{4n} \quad Q_n Q_n \mid Q_n^* \quad Q_{2n}^* \quad Q_{3n}^*$ | 6 |
| 89 | -7200 | 1 | $Q_{4n} \quad Q_n Q_n \mid \quad Q_{3n}^* \quad Q_{3n}^*$ | 6 |
| 90 | 8100 | 1 | $Q_{4n} \quad Q_n Q_n \mid Q_n^* Q_n^* \quad Q_{4n}^*$ | 6 |
| 91 | 180 | 1 | $Q_{4n} \quad Q_{2n} \quad \mid Q_n^* Q_n^* Q_n^* Q_n^* Q_n^* Q_n^*$ | 6 |
| 92 | -2700 | 1 | $Q_{4n} \quad Q_{2n} \quad \mid Q_n^* Q_n^* Q_n^* Q_n^* \quad Q_{2n}^*$ | 6 |
| 93 | 8100 | 1 | $Q_{4n} \quad Q_{2n} \quad \mid Q_n^* Q_n^* \quad Q_{2n}^* \quad Q_{2n}^*$ | 6 |
| 94 | -2700 | 1 | $Q_{4n} \quad Q_{2n} \quad \mid \quad Q_{2n}^* \quad Q_{2n}^* \quad Q_{2n}^*$ | 6 |
| 95 | 7200 | 1 | $Q_{4n} \quad Q_{2n} \quad \mid Q_n^* Q_n^* Q_n^* \quad Q_{3n}^*$ | 6 |
| 96 | -21600 | 1 | $Q_{4n} \quad Q_{2n} \quad \mid Q_n^* \quad Q_{2n}^* \quad Q_{3n}^*$ | 6 |
| 97 | 7200 | 1 | $Q_{4n} \quad Q_{2n} \quad \mid \quad Q_{3n}^* \quad Q_{3n}^*$ | 6 |
| 98 | -16200 | 1 | $Q_{4n} \quad Q_{2n} \quad \mid Q_n^* Q_n^* \quad Q_{4n}^*$ | 6 |
| 99 | 8100 | 1 | $Q_{4n} \quad Q_{2n} \quad \mid \quad Q_{2n}^* \quad Q_{4n}^*$ | 6 |
| 100 | 288 | 1 | $Q_{5n} \quad Q_n \mid Q_n^* Q_n^* Q_n^* Q_n^* Q_n^* Q_n^*$ | 6 |
| 101 | -4320 | 1 | $Q_{5n} \quad Q_n \mid Q_n^* Q_n^* Q_n^* Q_n^* \quad Q_{2n}^*$ | 6 |
| 102 | 12960 | 1 | $Q_{5n} \quad Q_n \mid Q_n^* Q_n^* \quad Q_{2n}^* \quad Q_{2n}^*$ | 6 |
| 103 | -4320 | 1 | $Q_{5n} \quad Q_n \mid \quad Q_{2n}^* \quad Q_{2n}^* \quad Q_{2n}^*$ | 6 |
| 104 | 11520 | 1 | $Q_{5n} \quad Q_n \mid Q_n^* Q_n^* Q_n^* \quad Q_{3n}^*$ | 6 |
| 105 | -34560 | 1 | $Q_{5n} \quad Q_n \mid Q_n^* \quad Q_{2n}^* \quad Q_{3n}^*$ | 6 |
| 106 | 11520 | 1 | $Q_{5n} \quad Q_n \mid \quad Q_{3n}^* \quad Q_{3n}^*$ | 6 |
| 107 | -25920 | 1 | $Q_{5n} \quad Q_n \mid Q_n^* Q_n^* \quad Q_{4n}^*$ | 6 |
| 108 | 25920 | 1 | $Q_{5n} \quad Q_n \mid Q_{2n}^* \quad Q_{4n}^*$ | 6 |
| 109 | 20736 | 1 | $Q_{5n} \quad Q_n \mid Q_n^* \quad Q_{5n}^*$ | 6 |
| 110 | -240 | 1 | $Q_{6n} \mid Q_n^* Q_n^* Q_n^* Q_n^* Q_n^* Q_n^*$ | 6 |
| 111 | 3600 | 1 | $Q_{6n} \mid Q_n^* Q_n^* Q_n^* Q_n^* \quad Q_{2n}^*$ | 6 |
| 112 | -10800 | 1 | $Q_{6n} \mid Q_n^* Q_n^* \quad Q_{2n}^* \quad Q_{2n}^*$ | 6 |
| 113 | 3600 | 1 | $Q_{6n} \mid Q_{2n}^* \quad Q_{2n}^* \quad Q_{2n}^*$ | 6 |
| 114 | -9600 | 1 | $Q_{6n} \mid Q_n^* Q_n^* Q_n^* \quad Q_{3n}^*$ | 6 |
| 115 | 28800 | 1 | $Q_{6n} \mid Q_n^* \quad Q_{2n}^* \quad Q_{3n}^*$ | 6 |
| 116 | -9600 | 1 | $Q_{6n} \mid Q_{3n}^* \quad Q_{3n}^*$ | 6 |
| 117 | 21600 | 1 | $Q_{6n} \mid Q_n^* Q_n^* \quad Q_{4n}^*$ | 6 |
| 118 | -21600 | 1 | $Q_{6n} \mid Q_{2n}^* \quad Q_{4n}^*$ | 6 |
| 119 | -34560 | 1 | $Q_{6n} \mid Q_n^* \quad Q_{5n}^*$ | 6 |
| 120 | 14400 | 1 | $Q_{6n} \mid Q_{6n}^*$ | 6 |

**Table 7** Coefficients, integer functions and basis vectors for calculation of the fourteen-particle azimuthal correlations.

| $i$ | $x_i$ | $f_i^{(m=7,l)}$ | **Basis vectors** | $l$ |
|---|---|---|---|---|
| 1 | -5040 | $M(M-8)(M-9)(M-10)(M-11)(M-12)(M-13)$ | 1 | 0 |
| 2 | 35280 | $(M-2)(M-9)(M-10)(M-11)(M-12)(M-13)$ | $Q_n \mid Q_n^*$ | 1 |
| 3 | -52920 | $(M-4)(M-10)(M-11)(M-12)(M-13)$ | $Q_n Q_n \mid Q_n^* Q_n^*$ | 2 |
| 4 | 105840 | $(M-4)(M-10)(M-11)(M-12)(M-13)$ | $Q_{2n} \mid Q_n^* Q_n^*$ | 2 |
| 5 | -52920 | $(M-4)(M-10)(M-11)(M-12)(M-13)$ | $Q_{2n} \mid Q_{2n}^*$ | 2 |
| 6 | 29400 | $(M-6)(M-11)(M-12)(M-13)$ | $Q_n Q_n Q_n \mid Q_n^* Q_n^* Q_n^*$ | 3 |
| 7 | -176400 | $(M-6)(M-11)(M-12)(M-13)$ | $Q_{2n} \quad Q_n \mid Q_n^* Q_n^* Q_n^*$ | 3 |
| 8 | 264600 | $(M-6)(M-11)(M-12)(M-13)$ | $Q_{2n} \quad Q_n \mid Q_n^* \quad Q_{2n}^*$ | 3 |
| 9 | 117600 | $(M-6)(M-11)(M-12)(M-13)$ | $Q_{3n} \mid Q_n^* Q_n^* Q_n^*$ | 3 |
| 10 | -352800 | $(M-6)(M-11)(M-12)(M-13)$ | $Q_{3n} \mid Q_n^* \quad Q_{2n}^*$ | 3 |
| 11 | 117600 | $(M-6)(M-11)(M-12)(M-13)$ | $Q_{3n} \mid Q_{3n}^*$ | 3 |
| 12 | -7350 | $(M-8)(M-12)(M-13)$ | $Q_n Q_n Q_n Q_n \mid Q_n^* Q_n^* Q_n^* Q_n^*$ | 4 |
| 13 | 88200 | $(M-8)(M-12)(M-13)$ | $Q_{2n} \quad Q_n Q_n \mid Q_n^* Q_n^* Q_n^* Q_n^*$ | 4 |
| 14 | -264600 | $(M-8)(M-12)(M-13)$ | $Q_{2n} \quad Q_n Q_n \mid Q_n^* Q_n^* \quad Q_{2n}^*$ | 4 |
| 15 | -44100 | $(M-8)(M-12)(M-13)$ | $Q_{2n} \quad Q_{2n} \mid Q_n^* Q_n^* Q_n^* Q_n^*$ | 4 |
| 16 | 264600 | $(M-8)(M-12)(M-13)$ | $Q_{2n} \quad Q_{2n} \mid Q_n^* Q_n^* \quad Q_{2n}^*$ | 4 |
| 17 | -66150 | $(M-8)(M-12)(M-13)$ | $Q_{2n} \quad Q_{2n} \mid Q_{2n}^* \quad Q_{2n}^*$ | 4 |
| 18 | -117600 | $(M-8)(M-12)(M-13)$ | $Q_{3n} \quad Q_n \mid Q_n^* Q_n^* Q_n^* Q_n^*$ | 4 |
| 19 | 705600 | $(M-8)(M-12)(M-13)$ | $Q_{3n}\quad Q_n \mid Q_n^* Q_n^*\; Q_{2n}^*$ | 4 |
| 20 | -352800 | $(M-8)(M-12)(M-13)$ | $Q_{3n}\quad Q_n \mid\; Q_{2n}^*\; Q_{2n}^*$ | 4 |
| 21 | -470400 | $(M-8)(M-12)(M-13)$ | $Q_{3n}\quad Q_n \mid Q_n^*\quad Q_{3n}^*$ | 4 |
| 22 | 88200 | $(M-8)(M-12)(M-13)$ | $Q_{4n}\qquad \mid Q_n^* Q_n^* Q_n^* Q_n^*$ | 4 |
| 23 | -529200 | $(M-8)(M-12)(M-13)$ | $Q_{4n}\qquad \mid Q_n^* Q_n^*\; Q_{2n}^*$ | 4 |
| 24 | 264600 | $(M-8)(M-12)(M-13)$ | $Q_{4n}\qquad \mid\; Q_{2n}^*\; Q_{2n}^*$ | 4 |
| 25 | 705600 | $(M-8)(M-12)(M-13)$ | $Q_{4n}\qquad \mid Q_n^*\quad Q_{3n}^*$ | 4 |
| 26 | -264600 | $(M-8)(M-12)(M-13)$ | $Q_{4n}\qquad \mid\qquad Q_{4n}^*$ | 4 |
| 27 | 882 | $(M-10)(M-13)$ | $Q_n Q_n Q_n Q_n Q_n \mid Q_n^* Q_n^* Q_n^* Q_n^* Q_n^*$ | 5 |
| 28 | -17640 | $(M-10)(M-13)$ | $Q_{2n}\; Q_n Q_n Q_n \mid Q_n^* Q_n^* Q_n^* Q_n^* Q_n^*$ | 5 |
| 29 | 88200 | $(M-10)(M-13)$ | $Q_{2n}\; Q_n Q_n Q_n \mid Q_n^* Q_n^* Q_n^*\; Q_{2n}^*$ | 5 |
| 30 | 26460 | $(M-10)(M-13)$ | $Q_{2n}\; Q_{2n}\; Q_n \mid Q_n^* Q_n^* Q_n^* Q_n^* Q_n^*$ | 5 |
| 31 | -264600 | $(M-10)(M-13)$ | $Q_{2n}\; Q_{2n}\; Q_n \mid Q_n^* Q_n^* Q_n^*\; Q_{2n}^*$ | 5 |
| 32 | 198450 | $(M-10)(M-13)$ | $Q_{2n}\; Q_{2n}\; Q_n \mid Q_n^*\; Q_{2n}^*\; Q_{2n}^*$ | 5 |
| 33 | 35280 | $(M-10)(M-13)$ | $Q_{3n}\quad Q_n Q_n \mid Q_n^* Q_n^* Q_n^* Q_n^* Q_n^*$ | 5 |
| 34 | -352800 | $(M-10)(M-13)$ | $Q_{3n}\quad Q_n Q_n \mid Q_n^* Q_n^* Q_n^*\; Q_{2n}^*$ | 5 |
| 35 | 529200 | $(M-10)(M-13)$ | $Q_{3n}\quad Q_n Q_n \mid Q_n^*\; Q_{2n}^*\; Q_{2n}^*$ | 5 |
| 36 | 352800 | $(M-10)(M-13)$ | $Q_{3n}\quad Q_n Q_n \mid Q_n^* Q_n^*\quad Q_{3n}^*$ | 5 |
| 37 | -35280 | $(M-10)(M-13)$ | $Q_{3n}\quad Q_{2n} \mid Q_n^* Q_n^* Q_n^* Q_n^* Q_n^*$ | 5 |
| 38 | 352800 | $(M-10)(M-13)$ | $Q_{3n}\quad Q_{2n} \mid Q_n^* Q_n^* Q_n^*\; Q_{2n}^*$ | 5 |
| 39 | -529200 | $(M-10)(M-13)$ | $Q_{3n}\quad Q_{2n} \mid Q_n^*\; Q_{2n}^*\; Q_{2n}^*$ | 5 |
| 40 | -705600 | $(M-10)(M-13)$ | $Q_{3n}\quad Q_{2n} \mid Q_n^* Q_n^*\quad Q_{3n}^*$ | 5 |
| 41 | 352800 | $(M-10)(M-13)$ | $Q_{3n}\quad Q_{2n} \mid\; Q_{2n}^*\; Q_{3n}^*$ | 5 |
| 42 | -52920 | $(M-10)(M-13)$ | $Q_{4n}\qquad Q_n \mid Q_n^* Q_n^* Q_n^* Q_n^* Q_n^*$ | 5 |
| 43 | 529200 | $(M-10)(M-13)$ | $Q_{4n}\qquad Q_n \mid Q_n^* Q_n^* Q_n^*\; Q_{2n}^*$ | 5 |
| 44 | -793800 | $(M-10)(M-13)$ | $Q_{4n}\qquad Q_n \mid Q_n^*\; Q_{2n}^*\; Q_{2n}^*$ | 5 |
| 45 | -1058400 | $(M-10)(M-13)$ | $Q_{4n}\qquad Q_n \mid Q_n^* Q_n^*\quad Q_{3n}^*$ | 5 |
| 46 | 1058400 | $(M-10)(M-13)$ | $Q_{4n}\qquad Q_n \mid\; Q_{2n}^*\; Q_{3n}^*$ | 5 |
| 47 | 793800 | $(M-10)(M-13)$ | $Q_{4n}\qquad Q_n \mid Q_n^*\quad Q_{4n}^*$ | 5 |
| 48 | 42336 | $(M-10)(M-13)$ | $Q_{5n}\qquad \mid Q_n^* Q_n^* Q_n^* Q_n^* Q_n^*$ | 5 |
| 49 | -423360 | $(M-10)(M-13)$ | $Q_{5n}\qquad \mid Q_n^* Q_n^* Q_n^*\; Q_{2n}^*$ | 5 |
| 50 | 635040 | $(M-10)(M-13)$ | $Q_{5n}\qquad \mid Q_n^*\; Q_{2n}^*\; Q_{2n}^*$ | 5 |
| 51 | 846720 | $(M-10)(M-13)$ | $Q_{5n}\qquad \mid Q_n^* Q_n^*\quad Q_{3n}^*$ | 5 |
| 52 | -846720 | $(M-10)(M-13)$ | $Q_{5n}\qquad \mid\; Q_{2n}^*\; Q_{3n}^*$ | 5 |
| 53 | -1270080 | $(M-10)(M-13)$ | $Q_{5n}\qquad \mid Q_n^*\quad Q_{4n}^*$ | 5 |
| 54 | 508032 | $(M-10)(M-13)$ | $Q_{5n} \mid Q^*_{5n}$ | 5 |
| 55 | -49 | $(M-12)$ | $Q_nQ_nQ_nQ_nQ_nQ_n \mid Q^*_nQ^*_nQ^*_nQ^*_nQ^*_nQ^*_n$ | 6 |
| 56 | 1470 | $(M-12)$ | $Q_{2n}\ Q_nQ_nQ_nQ_n \mid Q^*_nQ^*_nQ^*_nQ^*_nQ^*_nQ^*_n$ | 6 |
| 57 | -11025 | $(M-12)$ | $Q_{2n}\ Q_nQ_nQ_nQ_n \mid Q^*_nQ^*_nQ^*_nQ^*_n\ Q^*_{2n}$ | 6 |
| 58 | -4410 | $(M-12)$ | $Q_{2n}\ Q_{2n}\ Q_nQ_n \mid Q^*_nQ^*_nQ^*_nQ^*_nQ^*_nQ^*_n$ | 6 |
| 59 | 66150 | $(M-12)$ | $Q_{2n}\ Q_{2n}\ Q_nQ_n \mid Q^*_nQ^*_nQ^*_nQ^*_n\ Q^*_{2n}$ | 6 |
| 60 | -99225 | $(M-12)$ | $Q_{2n}\ Q_{2n}\ Q_nQ_n \mid Q^*_nQ^*_n\ Q^*_{2n}\ Q^*_{2n}$ | 6 |
| 61 | 1470 | $(M-12)$ | $Q_{2n}\ Q_{2n}\ Q_{2n} \mid Q^*_nQ^*_nQ^*_nQ^*_nQ^*_nQ^*_n$ | 6 |
| 62 | -22050 | $(M-12)$ | $Q_{2n}\ Q_{2n}\ Q_{2n} \mid Q^*_nQ^*_nQ^*_nQ^*_n\ Q^*_{2n}$ | 6 |
| 63 | 66150 | $(M-12)$ | $Q_{2n}\ Q_{2n}\ Q_{2n} \mid Q^*_nQ^*_n\ Q^*_{2n}\ Q^*_{2n}$ | 6 |
| 64 | -11025 | $(M-12)$ | $Q_{2n}\ Q_{2n}\ Q_{2n} \mid Q^*_{2n}\ Q^*_{2n}\ Q^*_{2n}$ | 6 |
| 65 | -3920 | $(M-12)$ | $Q_{3n}\ Q_nQ_nQ_n \mid Q^*_nQ^*_nQ^*_nQ^*_nQ^*_nQ^*_n$ | 6 |
| 66 | 58800 | $(M-12)$ | $Q_{3n}\ Q_nQ_nQ_n \mid Q^*_nQ^*_nQ^*_nQ^*_n\ Q^*_{2n}$ | 6 |
| 67 | -176400 | $(M-12)$ | $Q_{3n}\ Q_nQ_nQ_n \mid Q^*_nQ^*_n\ Q^*_{2n}\ Q^*_{2n}$ | 6 |
| 68 | 58800 | $(M-12)$ | $Q_{3n}\ Q_nQ_nQ_n \mid Q^*_{2n}\ Q^*_{2n}\ Q^*_{2n}$ | 6 |
| 69 | -78400 | $(M-12)$ | $Q_{3n}\ Q_nQ_nQ_n \mid Q^*_nQ^*_nQ^*_n\ Q^*_{3n}$ | 6 |
| 70 | 11760 | $(M-12)$ | $Q_{3n}\ Q_{2n}\ Q_n \mid Q^*_nQ^*_nQ^*_nQ^*_nQ^*_nQ^*_n$ | 6 |
| 71 | -176400 | $(M-12)$ | $Q_{3n}\ Q_{2n}\ Q_n \mid Q^*_nQ^*_nQ^*_nQ^*_n\ Q^*_{2n}$ | 6 |
| 72 | 529200 | $(M-12)$ | $Q_{3n}\ Q_{2n}\ Q_n \mid Q^*_nQ^*_n\ Q^*_{2n}\ Q^*_{2n}$ | 6 |
| 73 | -176400 | $(M-12)$ | $Q_{3n}\ Q_{2n}\ Q_n \mid Q^*_{2n}\ Q^*_{2n}\ Q^*_{2n}$ | 6 |
| 74 | 470400 | $(M-12)$ | $Q_{3n}\ Q_{2n}\ Q_n \mid Q^*_nQ^*_nQ^*_n\ Q^*_{3n}$ | 6 |
| 75 | -705600 | $(M-12)$ | $Q_{3n}\ Q_{2n}\ Q_n \mid Q^*_n\ Q^*_{2n}\ Q^*_{3n}$ | 6 |
| 76 | -3920 | $(M-12)$ | $Q_{3n}\ Q_{3n} \mid Q^*_nQ^*_nQ^*_nQ^*_nQ^*_nQ^*_n$ | 6 |
| 77 | 58800 | $(M-12)$ | $Q_{3n}\ Q_{3n} \mid Q^*_nQ^*_nQ^*_nQ^*_n\ Q^*_{2n}$ | 6 |
| 78 | -176400 | $(M-12)$ | $Q_{3n}\ Q_{3n} \mid Q^*_nQ^*_n\ Q^*_{2n}\ Q^*_{2n}$ | 6 |
| 79 | 58800 | $(M-12)$ | $Q_{3n}\ Q_{3n} \mid Q^*_{2n}\ Q^*_{2n}\ Q^*_{2n}$ | 6 |
| 80 | -156800 | $(M-12)$ | $Q_{3n}\ Q_{3n} \mid Q^*_nQ^*_nQ^*_n\ Q^*_{3n}$ | 6 |
| 81 | 470400 | $(M-12)$ | $Q_{3n}\ Q_{3n} \mid Q^*_n\ Q^*_{2n}\ Q^*_{3n}$ | 6 |
| 82 | -78400 | $(M-12)$ | $Q_{3n}\ Q_{3n} \mid Q^*_{3n}\ Q^*_{3n}$ | 6 |
| 83 | 8820 | $(M-12)$ | $Q_{4n}\ Q_nQ_n \mid Q^*_nQ^*_nQ^*_nQ^*_nQ^*_nQ^*_n$ | 6 |
| 84 | -132300 | $(M-12)$ | $Q_{4n}\ Q_nQ_n \mid Q^*_nQ^*_nQ^*_nQ^*_n\ Q^*_{2n}$ | 6 |
| 85 | 396900 | $(M-12)$ | $Q_{4n}\ Q_nQ_n \mid Q^*_nQ^*_n\ Q^*_{2n}\ Q^*_{2n}$ | 6 |
| 86 | -132300 | $(M-12)$ | $Q_{4n}\ Q_nQ_n \mid Q^*_{2n}\ Q^*_{2n}\ Q^*_{2n}$ | 6 |
| 87 | 352800 | $(M-12)$ | $Q_{4n}\ Q_nQ_n \mid Q^*_nQ^*_nQ^*_n\ Q^*_{3n}$ | 6 |
| 88 | -1058400 | $(M-12)$ | $Q_{4n}\ Q_nQ_n \mid Q^*_n\ Q^*_{2n}\ Q^*_{3n}$ | 6 |
| 89 | 352800 | $(M-12)$ | $Q_{4n}\ Q_nQ_n \mid Q^*_{3n}\ Q^*_{3n}$ | 6 |
| 90 | -396900 | $(M-12)$ | $Q_{4n}\ Q_nQ_n \mid Q^*_nQ^*_n\ Q^*_{4n}$ | 6 |
| 91 | -8820 | $(M-12)$ | $Q_{4n}\ Q_{2n} \mid Q^*_nQ^*_nQ^*_nQ^*_nQ^*_nQ^*_n$ | 6 |
| 92 | 132300 | $(M-12)$ | $Q_{4n}\ Q_{2n} \mid Q^*_nQ^*_nQ^*_nQ^*_n\ Q^*_{2n}$ | 6 |
| 93 | -396900 | $(M-12)$ | $Q_{4n} Q_{2n} \| Q_n^* Q_n^* Q_{2n}^* Q_{2n}^*$ | 6 |
| 94 | 132300 | $(M-12)$ | $Q_{4n} Q_{2n} \| Q_{2n}^* Q_{2n}^* Q_{2n}^*$ | 6 |
| 95 | -352800 | $(M-12)$ | $Q_{4n} Q_{2n} \| Q_n^* Q_n^* Q_n^* Q_{3n}^*$ | 6 |
| 96 | 1058400 | $(M-12)$ | $Q_{4n} Q_{2n} \| Q_n^* Q_{2n}^* Q_{3n}^*$ | 6 |
| 97 | -352800 | $(M-12)$ | $Q_{4n} Q_{2n} \| Q_{3n}^* Q_{3n}^*$ | 6 |
| 98 | 793800 | $(M-12)$ | $Q_{4n} Q_{2n} \| Q_n^* Q_n^* Q_{4n}^*$ | 6 |
| 99 | -396900 | $(M-12)$ | $Q_{4n} Q_{2n} \| Q_{2n}^* Q_{4n}^*$ | 6 |
| 100 | -14112 | $(M-12)$ | $Q_{5n} Q_n \| Q_n^* Q_n^* Q_n^* Q_n^* Q_n^* Q_n^*$ | 6 |
| 101 | 211680 | $(M-12)$ | $Q_{5n} Q_n \| Q_n^* Q_n^* Q_n^* Q_n^* Q_{2n}^*$ | 6 |
| 102 | -635040 | $(M-12)$ | $Q_{5n} Q_n \| Q_n^* Q_n^* Q_{2n}^* Q_{2n}^*$ | 6 |
| 103 | 211680 | $(M-12)$ | $Q_{5n} Q_n \| Q_{2n}^* Q_{2n}^* Q_{2n}^*$ | 6 |
| 104 | -564480 | $(M-12)$ | $Q_{5n} Q_n \| Q_n^* Q_n^* Q_n^* Q_{3n}^*$ | 6 |
| 105 | 1693440 | $(M-12)$ | $Q_{5n} Q_n \| Q_n^* Q_{2n}^* Q_{3n}^*$ | 6 |
| 106 | -564480 | $(M-12)$ | $Q_{5n} Q_n \| Q_{3n}^* Q_{3n}^*$ | 6 |
| 107 | 1270080 | $(M-12)$ | $Q_{5n} Q_n \| Q_n^* Q_n^* Q_{4n}^*$ | 6 |
| 108 | -1270080 | $(M-12)$ | $Q_{5n} Q_n \| Q_{2n}^* Q_{4n}^*$ | 6 |
| 109 | -1016064 | $(M-12)$ | $Q_{5n} Q_n \| Q_n^* Q_{5n}^*$ | 6 |
| 110 | 11760 | $(M-12)$ | $Q_{6n} \| Q_n^* Q_n^* Q_n^* Q_n^* Q_n^* Q_n^*$ | 6 |
| 111 | -176400 | $(M-12)$ | $Q_{6n} \| Q_n^* Q_n^* Q_n^* Q_n^* Q_{2n}^*$ | 6 |
| 112 | 529200 | $(M-12)$ | $Q_{6n} \| Q_n^* Q_n^* Q_{2n}^* Q_{2n}^*$ | 6 |
| 113 | -176400 | $(M-12)$ | $Q_{6n} \| Q_{2n}^* Q_{2n}^* Q_{2n}^*$ | 6 |
| 114 | 470400 | $(M-12)$ | $Q_{6n} \| Q_n^* Q_n^* Q_n^* Q_{3n}^*$ | 6 |
| 115 | -1411200 | $(M-12)$ | $Q_{6n} \| Q_n^* Q_{2n}^* Q_{3n}^*$ | 6 |
| 116 | 470400 | $(M-12)$ | $Q_{6n} \| Q_{3n}^* Q_{3n}^*$ | 6 |
| 117 | -1058400 | $(M-12)$ | $Q_{6n} \| Q_n^* Q_n^* Q_{4n}^*$ | 6 |
| 118 | 1058400 | $(M-12)$ | $Q_{6n} \| Q_{2n}^* Q_{4n}^*$ | 6 |
| 119 | 1693440 | $(M-12)$ | $Q_{6n} \| Q_n^* Q_{5n}^*$ | 6 |
| 120 | -705600 | $(M-12)$ | $Q_{6n} \| Q_{6n}^*$ | 6 |
| 121 | 1 | 1 | $Q_n Q_n Q_n Q_n Q_n Q_n Q_n \| Q_n^* Q_n^* Q_n^* Q_n^* Q_n^* Q_n^* Q_n^*$ | 7 |
| 122 | -42 | 1 | $Q_{2n} Q_n Q_n Q_n Q_n Q_n \| Q_n^* Q_n^* Q_n^* Q_n^* Q_n^* Q_n^* Q_n^*$ | 7 |
| 123 | 441 | 1 | $Q_{2n} Q_n Q_n Q_n Q_n Q_n \| Q_n^* Q_n^* Q_n^* Q_n^* Q_n^* Q_{2n}^*$ | 7 |
| 124 | 210 | 1 | $Q_{2n} Q_{2n} Q_n Q_n Q_n \| Q_n^* Q_n^* Q_n^* Q_n^* Q_n^* Q_n^* Q_n^*$ | 7 |
| 125 | -4410 | 1 | $Q_{2n} Q_{2n} Q_n Q_n Q_n \| Q_n^* Q_n^* Q_n^* Q_n^* Q_n^* Q_{2n}^*$ | 7 |
| 126 | 11025 | 1 | $Q_{2n} Q_{2n} Q_n Q_n Q_n \| Q_n^* Q_n^* Q_n^* Q_{2n}^* Q_{2n}^*$ | 7 |
| 127 | -210 | 1 | $Q_{2n} Q_{2n} Q_{2n} Q_n \| Q_n^* Q_n^* Q_n^* Q_n^* Q_n^* Q_n^* Q_n^*$ | 7 |
| 128 | 4410 | 1 | $Q_{2n} Q_{2n} Q_{2n} Q_n \| Q_n^* Q_n^* Q_n^* Q_n^* Q_n^* Q_{2n}^*$ | 7 |
| 129 | -22050 | 1 | $Q_{2n} Q_{2n} Q_{2n} Q_n \| Q_n^* Q_n^* Q_n^* Q_{2n}^* Q_{2n}^*$ | 7 |
| 130 | 11025 | 1 | $Q_{2n} Q_{2n} Q_{2n} Q_n \| Q_n^* Q_{2n}^* Q_{2n}^* Q_{2n}^*$ | 7 |
| 131 | 140 | 1 | $Q_{3n} Q_n Q_n Q_n Q_n \| Q_n^* Q_n^* Q_n^* Q_n^* Q_n^* Q_n^* Q_n^*$ | 7 |
| 132 | -2940 | 1 | $Q_{3n} Q_n Q_n Q_n Q_n \| Q_n^* Q_n^* Q_n^* Q_n^* Q_n^* Q_{2n}^*$ | 7 |
| 133 | 14700 | 1 | $Q_{3n} \quad Q_nQ_nQ_nQ_n \,\vert\, Q_n^*Q_n^*Q_n^* \; Q_{2n}^* \; Q_{2n}^*$ | 7 |
| 134 | -14700 | 1 | $Q_{3n} \quad Q_nQ_nQ_nQ_n \,\vert\, Q_n^* \; Q_{2n}^* \; Q_{2n}^* \; Q_{2n}^*$ | 7 |
| 135 | 4900 | 1 | $Q_{3n} \quad Q_nQ_nQ_nQ_n \,\vert\, Q_n^*Q_n^*Q_n^*Q_n^* \; Q_{3n}^*$ | 7 |
| 136 | -840 | 1 | $Q_{3n} \quad Q_{2n} \; Q_nQ_n \,\vert\, Q_n^*Q_n^*Q_n^*Q_n^*Q_n^*Q_n^*Q_n^*$ | 7 |
| 137 | 17640 | 1 | $Q_{3n} \quad Q_{2n} \; Q_nQ_n \,\vert\, Q_n^*Q_n^*Q_n^*Q_n^*Q_n^* \; Q_{2n}^*$ | 7 |
| 138 | -88200 | 1 | $Q_{3n} \quad Q_{2n} \; Q_nQ_n \,\vert\, Q_n^*Q_n^*Q_n^* \; Q_{2n}^* \; Q_{2n}^*$ | 7 |
| 139 | 88200 | 1 | $Q_{3n} \quad Q_{2n} \; Q_nQ_n \,\vert\, Q_n^* \; Q_{2n}^* \; Q_{2n}^* \; Q_{2n}^*$ | 7 |
| 140 | -58800 | 1 | $Q_{3n} \quad Q_{2n} \; Q_nQ_n \,\vert\, Q_n^*Q_n^*Q_n^*Q_n^* \; Q_{3n}^*$ | 7 |
| 141 | 176400 | 1 | $Q_{3n} \quad Q_{2n} \; Q_nQ_n \,\vert\, Q_n^*Q_n^* \; Q_{2n}^* \; Q_{3n}^*$ | 7 |
| 142 | 420 | 1 | $Q_{3n} \quad Q_{2n} \; Q_{2n} \,\vert\, Q_n^*Q_n^*Q_n^*Q_n^*Q_n^*Q_n^*Q_n^*$ | 7 |
| 143 | -8820 | 1 | $Q_{3n} \quad Q_{2n} \; Q_{2n} \,\vert\, Q_n^*Q_n^*Q_n^*Q_n^*Q_n^* \; Q_{2n}^*$ | 7 |
| 144 | 44100 | 1 | $Q_{3n} \quad Q_{2n} \; Q_{2n} \,\vert\, Q_n^*Q_n^*Q_n^* \; Q_{2n}^* \; Q_{2n}^*$ | 7 |
| 145 | -44100 | 1 | $Q_{3n} \quad Q_{2n} \; Q_{2n} \,\vert\, Q_n^* \; Q_{2n}^* \; Q_{2n}^* \; Q_{2n}^*$ | 7 |
| 146 | 29400 | 1 | $Q_{3n} \quad Q_{2n} \; Q_{2n} \,\vert\, Q_n^*Q_n^*Q_n^*Q_n^* \; Q_{3n}^*$ | 7 |
| 147 | -176400 | 1 | $Q_{3n} \quad Q_{2n} \; Q_{2n} \,\vert\, Q_n^*Q_n^* \; Q_{2n}^* \; Q_{3n}^*$ | 7 |
| 148 | 44100 | 1 | $Q_{3n} \quad Q_{2n} \; Q_{2n} \,\vert\, Q_{2n}^* \; Q_{2n}^* \; Q_{3n}^*$ | 7 |
| 149 | 560 | 1 | $Q_{3n} \quad Q_{3n} \; Q_n \,\vert\, Q_n^*Q_n^*Q_n^*Q_n^*Q_n^*Q_n^*Q_n^*$ | 7 |
| 150 | -11760 | 1 | $Q_{3n} \quad Q_{3n} \; Q_n \,\vert\, Q_n^*Q_n^*Q_n^*Q_n^*Q_n^* \; Q_{2n}^*$ | 7 |
| 151 | 58800 | 1 | $Q_{3n} \quad Q_{3n} \; Q_n \,\vert\, Q_n^*Q_n^*Q_n^* \; Q_{2n}^* \; Q_{2n}^*$ | 7 |
| 152 | -58800 | 1 | $Q_{3n} \quad Q_{3n} \; Q_n \,\vert\, Q_n^* \; Q_{2n}^* \; Q_{2n}^* \; Q_{2n}^*$ | 7 |
| 153 | 39200 | 1 | $Q_{3n} \quad Q_{3n} \; Q_n \,\vert\, Q_n^*Q_n^*Q_n^*Q_n^* \; Q_{3n}^*$ | 7 |
| 154 | -235200 | 1 | $Q_{3n} \quad Q_{3n} \; Q_n \,\vert\, Q_n^*Q_n^* \; Q_{2n}^* \; Q_{3n}^*$ | 7 |
| 155 | 117600 | 1 | $Q_{3n} \quad Q_{3n} \; Q_n \,\vert\, Q_{2n}^* \; Q_{2n}^* \; Q_{3n}^*$ | 7 |
| 156 | 78400 | 1 | $Q_{3n} \quad Q_{3n} \; Q_n \,\vert\, Q_n^* \; Q_{3n}^* \; Q_{3n}^*$ | 7 |
| 157 | -420 | 1 | $Q_{4n} \quad Q_nQ_nQ_n \,\vert\, Q_n^*Q_n^*Q_n^*Q_n^*Q_n^*Q_n^*Q_n^*$ | 7 |
| 158 | 8820 | 1 | $Q_{4n} \quad Q_nQ_nQ_n \,\vert\, Q_n^*Q_n^*Q_n^*Q_n^*Q_n^* \; Q_{2n}^*$ | 7 |
| 159 | -44100 | 1 | $Q_{4n} \quad Q_nQ_nQ_n \,\vert\, Q_n^*Q_n^*Q_n^* \; Q_{2n}^* \; Q_{2n}^*$ | 7 |
| 160 | 44100 | 1 | $Q_{4n} \quad Q_nQ_nQ_n \,\vert\, Q_n^* \; Q_{2n}^* \; Q_{2n}^* \; Q_{2n}^*$ | 7 |
| 161 | -29400 | 1 | $Q_{4n} \quad Q_nQ_nQ_n \,\vert\, Q_n^*Q_n^*Q_n^*Q_n^* \; Q_{3n}^*$ | 7 |
| 162 | 176400 | 1 | $Q_{4n} \quad Q_nQ_nQ_n \,\vert\, Q_n^*Q_n^* \; Q_{2n}^* \; Q_{3n}^*$ | 7 |
| 163 | -88200 | 1 | $Q_{4n} \quad Q_nQ_nQ_n \,\vert\, Q_{2n}^* \; Q_{2n}^* \; Q_{3n}^*$ | 7 |
| 164 | -117600 | 1 | $Q_{4n} \quad Q_nQ_nQ_n \,\vert\, Q_n^* \; Q_{3n}^* \; Q_{3n}^*$ | 7 |
| 165 | 44100 | 1 | $Q_{4n} \quad Q_nQ_nQ_n \,\vert\, Q_n^*Q_n^*Q_n^* \; Q_{4n}^*$ | 7 |
| 166 | 1260 | 1 | $Q_{4n} \quad Q_{2n} \; Q_n \,\vert\, Q_n^*Q_n^*Q_n^*Q_n^*Q_n^*Q_n^*Q_n^*$ | 7 |
| 167 | -26460 | 1 | $Q_{4n} \quad Q_{2n} \; Q_n \,\vert\, Q_n^*Q_n^*Q_n^*Q_n^*Q_n^* \; Q_{2n}^*$ | 7 |
| 168 | 132300 | 1 | $Q_{4n} \quad Q_{2n} \; Q_n \,\vert\, Q_n^*Q_n^*Q_n^* \; Q_{2n}^* \; Q_{2n}^*$ | 7 |
| 169 | -132300 | 1 | $Q_{4n} \quad Q_{2n} \; Q_n \,\vert\, Q_n^* \; Q_{2n}^* \; Q_{2n}^* \; Q_{2n}^*$ | 7 |
| 170 | 88200 | 1 | $Q_{4n} \quad Q_{2n} \; Q_n \,\vert\, Q_n^*Q_n^*Q_n^*Q_n^* \; Q_{3n}^*$ | 7 |
| 171 | -529200 | 1 | $Q_{4n} \quad Q_{2n} \; Q_n \,\vert\, Q_n^*Q_n^* \; Q_{2n}^* \; Q_{3n}^*$ | 7 |
| 172 | 264600 | 1 | $Q_{4n} \quad Q_{2n} \; Q_n \,\vert\, Q_{2n}^* \; Q_{2n}^* \; Q_{3n}^*$ | 7 |
| 173 | 352800 | 1 | $Q_{4n} Q_{2n} Q_n \mid Q_n^* Q_{3n}^* Q_{3n}^*$ | 7 |
| 174 | -264600 | 1 | $Q_{4n} Q_{2n} Q_n \mid Q_n^* Q_n^* Q_n^* Q_{4n}^*$ | 7 |
| 175 | 396900 | 1 | $Q_{4n} Q_{2n} Q_n \mid Q_n^* Q_{2n}^* Q_{4n}^*$ | 7 |
| 176 | -840 | 1 | $Q_{4n} Q_{3n} \mid Q_n^* Q_n^* Q_n^* Q_n^* Q_n^* Q_n^* Q_n^*$ | 7 |
| 177 | 17640 | 1 | $Q_{4n} Q_{3n} \mid Q_n^* Q_n^* Q_n^* Q_n^* Q_n^* Q_{2n}^*$ | 7 |
| 178 | -88200 | 1 | $Q_{4n} Q_{3n} \mid Q_n^* Q_n^* Q_n^* Q_{2n}^* Q_{2n}^*$ | 7 |
| 179 | 88200 | 1 | $Q_{4n} Q_{3n} \mid Q_n^* Q_{2n}^* Q_{2n}^* Q_{2n}^*$ | 7 |
| 180 | -58800 | 1 | $Q_{4n} Q_{3n} \mid Q_n^* Q_n^* Q_n^* Q_n^* Q_{3n}^*$ | 7 |
| 181 | 352800 | 1 | $Q_{4n} Q_{3n} \mid Q_n^* Q_n^* Q_{2n}^* Q_{3n}^*$ | 7 |
| 182 | -176400 | 1 | $Q_{4n} Q_{3n} \mid Q_{2n}^* Q_{2n}^* Q_{3n}^*$ | 7 |
| 183 | -235200 | 1 | $Q_{4n} Q_{3n} \mid Q_n^* Q_{3n}^* Q_{3n}^*$ | 7 |
| 184 | 176400 | 1 | $Q_{4n} Q_{3n} \mid Q_n^* Q_n^* Q_n^* Q_{4n}^*$ | 7 |
| 185 | -529200 | 1 | $Q_{4n} Q_{3n} \mid Q_n^* Q_{2n}^* Q_{4n}^*$ | 7 |
| 186 | 176400 | 1 | $Q_{4n} Q_{3n} \mid Q_{3n}^* Q_{4n}^*$ | 7 |
| 187 | 1008 | 1 | $Q_{5n} Q_n Q_n \mid Q_n^* Q_n^* Q_n^* Q_n^* Q_n^* Q_n^* Q_n^*$ | 7 |
| 188 | -21168 | 1 | $Q_{5n} Q_n Q_n \mid Q_n^* Q_n^* Q_n^* Q_n^* Q_n^* Q_{2n}^*$ | 7 |
| 189 | 105840 | 1 | $Q_{5n} Q_n Q_n \mid Q_n^* Q_n^* Q_n^* Q_{2n}^* Q_{2n}^*$ | 7 |
| 190 | -105840 | 1 | $Q_{5n} Q_n Q_n \mid Q_n^* Q_{2n}^* Q_{2n}^* Q_{2n}^*$ | 7 |
| 191 | 70560 | 1 | $Q_{5n} Q_n Q_n \mid Q_n^* Q_n^* Q_n^* Q_n^* Q_{3n}^*$ | 7 |
| 192 | -423360 | 1 | $Q_{5n} Q_n Q_n \mid Q_n^* Q_n^* Q_{2n}^* Q_{3n}^*$ | 7 |
| 193 | 211680 | 1 | $Q_{5n} Q_n Q_n \mid Q_{2n}^* Q_{2n}^* Q_{3n}^*$ | 7 |
| 194 | 282240 | 1 | $Q_{5n} Q_n Q_n \mid Q_n^* Q_{3n}^* Q_{3n}^*$ | 7 |
| 195 | -211680 | 1 | $Q_{5n} Q_n Q_n \mid Q_n^* Q_n^* Q_n^* Q_{4n}^*$ | 7 |
| 196 | 635040 | 1 | $Q_{5n} Q_n Q_n \mid Q_n^* Q_{2n}^* Q_{4n}^*$ | 7 |
| 197 | -423360 | 1 | $Q_{5n} Q_n Q_n \mid Q_{3n}^* Q_{4n}^*$ | 7 |
| 198 | 254016 | 1 | $Q_{5n} Q_n Q_n \mid Q_n^* Q_n^* Q_{5n}^*$ | 7 |
| 199 | -1008 | 1 | $Q_{5n} Q_{2n} \mid Q_n^* Q_n^* Q_n^* Q_n^* Q_n^* Q_n^* Q_n^*$ | 7 |
| 200 | 21168 | 1 | $Q_{5n} Q_{2n} \mid Q_n^* Q_n^* Q_n^* Q_n^* Q_n^* Q_{2n}^*$ | 7 |
| 201 | -105840 | 1 | $Q_{5n} Q_{2n} \mid Q_n^* Q_n^* Q_n^* Q_{2n}^* Q_{2n}^*$ | 7 |
| 202 | 105840 | 1 | $Q_{5n} Q_{2n} \mid Q_n^* Q_{2n}^* Q_{2n}^* Q_{2n}^*$ | 7 |
| 203 | -70560 | 1 | $Q_{5n} Q_{2n} \mid Q_n^* Q_n^* Q_n^* Q_n^* Q_{3n}^*$ | 7 |
| 204 | 423360 | 1 | $Q_{5n} Q_{2n} \mid Q_n^* Q_n^* Q_{2n}^* Q_{3n}^*$ | 7 |
| 205 | -211680 | 1 | $Q_{5n} Q_{2n} \mid Q_{2n}^* Q_{2n}^* Q_{3n}^*$ | 7 |
| 206 | -282240 | 1 | $Q_{5n} Q_{2n} \mid Q_n^* Q_{3n}^* Q_{3n}^*$ | 7 |
| 207 | 211680 | 1 | $Q_{5n} Q_{2n} \mid Q_n^* Q_n^* Q_n^* Q_{4n}^*$ | 7 |
| 208 | -635040 | 1 | $Q_{5n} Q_{2n} \mid Q_n^* Q_{2n}^* Q_{4n}^*$ | 7 |
| 209 | 423360 | 1 | $Q_{5n} Q_{2n} \mid Q_{3n}^* Q_{4n}^*$ | 7 |
| 210 | -508032 | 1 | $Q_{5n} Q_{2n} \mid Q_n^* Q_n^* Q_{5n}^*$ | 7 |
| 211 | 254016 | 1 | $Q_{5n} Q_{2n} \mid Q_{2n}^* Q_{5n}^*$ | 7 |
| 212 | -1680 | 1 | $Q_{6n} Q_n \mid Q_n^* Q_n^* Q_n^* Q_n^* Q_n^* Q_n^* Q_n^*$ | 7 |
| 213 | 35280 | 1 | $Q_{6n} \quad Q_n \mid Q_n^* Q_n^* Q_n^* Q_n^* Q_n^* Q_{2n}^*$ | 7 |
| 214 | -176400 | 1 | $Q_{6n} \quad Q_n \mid Q_n^* Q_n^* Q_n^* Q_{2n}^* Q_{2n}^*$ | 7 |
| 215 | 176400 | 1 | $Q_{6n} \quad Q_n \mid Q_n^* Q_{2n}^* Q_{2n}^* Q_{2n}^*$ | 7 |
| 216 | -117600 | 1 | $Q_{6n} \quad Q_n \mid Q_n^* Q_n^* Q_n^* Q_n^* Q_{3n}^*$ | 7 |
| 217 | 705600 | 1 | $Q_{6n} \quad Q_n \mid Q_n^* Q_n^* Q_{2n}^* Q_{3n}^*$ | 7 |
| 218 | -352800 | 1 | $Q_{6n} \quad Q_n \mid Q_{2n}^* Q_{2n}^* Q_{3n}^*$ | 7 |
| 219 | -470400 | 1 | $Q_{6n} \quad Q_n \mid Q_n^* Q_{3n}^* Q_{3n}^*$ | 7 |
| 220 | 352800 | 1 | $Q_{6n} \quad Q_n \mid Q_n^* Q_n^* Q_n^* Q_{4n}^*$ | 7 |
| 221 | -1058400 | 1 | $Q_{6n} \quad Q_n \mid Q_n^* Q_{2n}^* Q_{4n}^*$ | 7 |
| 222 | 705600 | 1 | $Q_{6n} \quad Q_n \mid Q_{3n}^* Q_{4n}^*$ | 7 |
| 223 | -846720 | 1 | $Q_{6n} \quad Q_n \mid Q_n^* Q_n^* Q_{5n}^*$ | 7 |
| 224 | 846720 | 1 | $Q_{6n} \quad Q_n \mid Q_{2n}^* Q_{5n}^*$ | 7 |
| 225 | 705600 | 1 | $Q_{6n} \quad Q_n \mid Q_n^* Q_{6n}^*$ | 7 |
| 226 | 1440 | 1 | $Q_{7n} \mid Q_n^* Q_n^* Q_n^* Q_n^* Q_n^* Q_n^* Q_n^*$ | 7 |
| 227 | -30240 | 1 | $Q_{7n} \mid Q_n^* Q_n^* Q_n^* Q_n^* Q_n^* Q_{2n}^*$ | 7 |
| 228 | 151200 | 1 | $Q_{7n} \mid Q_n^* Q_n^* Q_n^* Q_{2n}^* Q_{2n}^*$ | 7 |
| 229 | -151200 | 1 | $Q_{7n} \mid Q_n^* Q_{2n}^* Q_{2n}^* Q_{2n}^*$ | 7 |
| 230 | 100800 | 1 | $Q_{7n} \mid Q_n^* Q_n^* Q_n^* Q_n^* Q_{3n}^*$ | 7 |
| 231 | -604800 | 1 | $Q_{7n} \mid Q_n^* Q_n^* Q_{2n}^* Q_{3n}^*$ | 7 |
| 232 | 302400 | 1 | $Q_{7n} \mid Q_{2n}^* Q_{2n}^* Q_{3n}^*$ | 7 |
| 233 | 403200 | 1 | $Q_{7n} \mid Q_n^* Q_{3n}^* Q_{3n}^*$ | 7 |
| 234 | -302400 | 1 | $Q_{7n} \mid Q_n^* Q_n^* Q_n^* Q_{4n}^*$ | 7 |
| 235 | 907200 | 1 | $Q_{7n} \mid Q_n^* Q_{2n}^* Q_{4n}^*$ | 7 |
| 236 | -604800 | 1 | $Q_{7n} \mid Q_{3n}^* Q_{4n}^*$ | 7 |
| 237 | 725760 | 1 | $Q_{7n} \mid Q_n^* Q_n^* Q_{5n}^*$ | 7 |
| 238 | -725760 | 1 | $Q_{7n} \mid Q_{2n}^* Q_{5n}^*$ | 7 |
| 239 | -1209600 | 1 | $Q_{7n} \mid Q_n^* Q_{6n}^*$ | 7 |
| 240 | 518400 | 1 | $Q_{7n} \mid Q_{7n}^*$ | 7 |